\documentclass[fleqn,usenatbib]{mnras}
\usepackage{newtxtext,newtxmath}
\usepackage[T1]{fontenc}
\DeclareRobustCommand{\VAN}[3]{#2}
\let\VANthebibliography\thebibliography
\def\thebibliography{\DeclareRobustCommand{\VAN}[3]{##3}\VANthebibliography}

\usepackage{graphicx}	% Including figure files
\usepackage{color}
\usepackage{amsmath}	% Advanced maths commands

\usepackage{multirow}
\usepackage{braket}
\usepackage{comment}
\usepackage{bm}
\usepackage{graphicx}
\usepackage{soul}

\usepackage[dvipsnames]{xcolor}
\usepackage[capitalise]{cleveref}
\crefformat{equation}{Eq. (#2#1#3)}
\crefrangeformat{equation}{Eqs. #3#1#4-#5#2#6}
\crefrangeformat{figure}{Figs. #3#1#4-#5#2#6}
\Crefformat{equation}{Equation (#2#1#3)}
\Crefname{equation}{Equation}{Equations}

\title[CosmoGLINT: Transformer-Based LIM Generator]{CosmoGLINT: Cosmological Generative Model for Line Intensity Mapping with Transformer}

\author[K. Moriwaki et al.]{
Kana Moriwaki,$^{1,2,3}$\thanks{E-mail: kana.moriwaki@phys.s.u-tokyo.ac.jp}, 
Rui Lan Jun,$^{2}$
Ken Osato,$^{4,5,3,6}$
Naoki Yoshida$^{2,3,6}$
\\
$^{1}$Research Center for the Early Universe, Graduate School of Science, The University of Tokyo, 7-3-1 Hongo, Bunkyo, Tokyo 113-0033, Japan\\
$^{2}$Department of Physics, Graduate School of Science, The University of Tokyo, 7-3-1 Hongo, Bunkyo, Tokyo 133-0033, Japan\\
$^{3}$RIKEN Center for Advanced Intelligence Project, 1-4-1 Nihonbashi, Chuo, Tokyo 103-0027, Japan\\
$^{4}$Center for Frontier Science, Chiba University, 1-33 Yayoi-cho, Inage-ku, Chiba 263-8522, Japan\\
$^{5}$Department of Physics, Graduate School of Science, Chiba University, 1-33 Yayoi-cho, Inage-ku, Chiba 263-8522, Japan\\
$^{6}$Kavli Institute for the Physics and Mathematics of the Universe, The University of Tokyo Institutes for Advanced Study,\\
5-1-5 Kashiwanoha, Kashiwa-shi, Chiba, 277-8583, Japan\\
}

\date{Accepted XXX. Received YYY; in original form ZZZ}
\pubyear{2025}

\begin{document}
\label{firstpage}
\pagerange{\pageref{firstpage}--\pageref{lastpage}}
\maketitle

\begin{abstract}
Modelling star-forming galaxies is crucial for upcoming observations of large-scale matter and galaxy distributions with galaxy redshift surveys and line intensity mapping (LIM).
We introduce CosmoGLINT (Cosmological Generative model for Line INtensity mapping with Transformer), a Transformer-based generative framework designed to create realistic galaxy populations from dark matter (DM)-only simulations.
CosmoGLINT auto-regressively generates sequences of galaxy properties --- including star formation rate (SFR), distance to the halo centre, and radial and tangential velocities relative to the halo --- conditioned on halo mass.
Trained on the IllustrisTNG hydrodynamic simulation, the model reproduces key statistical properties of the original data, including the voxel intensity distribution and the power spectrum both in real and redshift space. 
It can efficiently generate a number of different realisations of the designated galaxy populations, enabling the creation of mock LIM/redshift survey catalogues from large halo catalogues produced by fast DM-only simulations.
We show that our model, trained at multiple redshifts, can be applied to DM halo lightcone data to generate a realistic mock galaxy lightcone that incorporates the redshift evolution
of the galaxy population.
The mock catalogues can be readily used to derive statistical quantities and to develop data analysis pipelines for ongoing and future wide-field surveys.
\end{abstract}


%%%%%%%%%%%%%%%%% BODY OF PAPER %%%%%%%%%%%%%%%%%%

\section{Introduction}

The large-scale structure (LSS) of the Universe is a powerful probe of cosmology and galaxy formation.
Line intensity mapping (LIM) is a promising technique for probing the LSS by collecting the cumulative line emissions from galaxies \citep[see][for a review]{Kovetz17}.
Previous studies largely focused on the 21cm line from neutral hydrogen \citep[e.g.,][]{Chang10}, and there are growing interests in other emission lines such as Ly$\alpha$, H$\alpha$, COs, and [\ion{C}{ii}], which primarily originate from nebular regions of star-forming galaxies. Several observations have already been conducted for some of these lines \citep{Keating16, Keating20, Lunde24}, and further large-scale observations are ongoing or planned \citep[e.g.,][]{Crites14, Aguirre18, Concerto20, EXCLAIM20, Crill20_SPHEREx, CCAT-Prime23}. 

It is essential to use realistic mock observational catalogues for preparing and analysing LIM data: mock data can be used to predict observable signals, estimate the statistical covariance of LIM statistics, guide the design of observational strategies, or validate data analysis pipelines \citep[e.g.,][]{Lunde24}. The mock data can also be used to develop methodologies for analysing observational data. For instance, \citet{Moriwaki20} proposed to use a machine learning model to remove line interlopers from noisy observational maps, for which more than 100 independent mock intensity maps are required. 

One may naively hope to directly use the outputs of hydrodynamics simulations, which numerically solve the gas dynamics and therefore model the star formation in a physically motivated manner, to create realistic mocks \citep[e.g.,][]{Karoumpis22}. Unfortunately, hydrodynamic simulations are computationally expensive and thus it is intractable to generate a number of mocks for statistical analysis. 
Another option is to adopt a heuristic relation between dark matter (DM) and baryonic components to {\it paint} the output of DM-only simulations \citep{Yue15, Silva15, Bethermin22, Gkogkou23, Roy23, Mas-Ribas23}. In our previous studies \citep{Moriwaki20, Moriwaki21a, Moriwaki21b}, for instance, we fitted the relation between halo mass and total luminosity of its member galaxies in a hydrodynamics simulation, and assigned luminosities to DM haloes accordingly. 
In such approaches, it is common to assign a single value of luminosity to each halo.
While this simplification is reasonable when the observation does not resolve individual member galaxies or when every halo has only one member galaxy, this is not the case for many LIM observations.\footnote{For instance, SPHEREx \citep{Dore18,Crill20_SPHEREx} has 6.2 arcsec resolution, which corresponds to 0.16 Mpc at $z = 2$. The virial radius of a halo with $M = 10^{13}~\rm M_\odot$ ($10^{14}~\rm M_\odot$) at this redshift is 0.22 Mpc (0.47 Mpc), and the emission-line galaxies are considered to be more widely distributed, typically extending several times beyond this radius \citep{Orsi18,Avila20,Jun25}. The effective resolution in comoving scale is larger at lower redshift.} 
The individual galaxy distribution should be taken into account when performing pixel-level analyses --- such as masking, analysis of voxel intensity distribution, and machine-learning image processing --- as well as analysis of the small-scale power spectrum \citep{Jun25}.

In the context of conventional galaxy surveys, the halo occupation distribution model \citep[HOD;][]{Berlind02}, which models the probability distribution of the number of galaxies $N$ that satisfy observational selection criteria for a given halo mass, $P(N|M)$, is commonly used \citep[e.g.,][]{DESI_HOD}.
For LIM experiments, which measure intensity distribution rather than number density distribution, it is necessary to model not only the number of galaxies per halo but also the luminosities of individual galaxies as 
\begin{align}
    \label{eq:L_prob}
    p(L_0, L_1, \cdots , L_N | M).
\end{align}
Since the properties of member galaxies are often correlated --- the so-called galactic conformity\footnote{
Observations have shown that quenched central galaxies tend to be surrounded by a higher fraction of quenched satellites, both in the local universe \citep[e.g.,][]{Weinmann06} and even at $z = 2 \text{--} 3$ \citep[e.g.,][]{McConachie25}. This phenomenon is called galactic conformity.
Galactic conformity has also been observed in hydrodynamic simulations, including TNG \citep{Ayromlou23}.
Several physical mechanisms have been proposed to explain this phenomenon, such as assembly bias \citep{Hearin16}, environmental effects \citep{Sun18}, and AGN feedback \citep{Kauffmann15}, although it remains unclear which is primarily responsible for the phenomenon.
While it has been suggested that the correlation between  properties of nearby galaxies could exist even when the separation is larger than the halo size \citep[e.g.,][]{Kauffmann13}, in this study we refer to galactic conformity as the correlation of galaxy properties within the same halo for simplicity.
} --- and such correlations affect LIM observables \citep{Jun25b}, it is crucial to model their joint probability distribution, as in \cref{eq:L_prob}, instead of modelling them independently.

Accurate modelling of halo-galaxy and galaxy-galaxy connections would not only benefit LIM surveys but also allow us to fully exploit the information from other observations such as local satellite galaxy surveys \citep[e.g.,][]{Carlsten22,Mao24} and other large-scale cosmological surveys \citep[e.g.,][]{Takada14, DESI16, Euclid25, LSST09, Roman15} for understanding cosmology and galaxy formation and evolution.

Recently, HOD models that account for galactic conformity have been proposed for conventional galaxy redshift surveys \citep{Reyes-Peraza24,Yuan25}. These models condition the mean number of detectable satellites on the presence of a detectable central galaxy within a halo.
However, constructing such a model for LIM, i.e., for the probability distribution in a high-dimensional space (\cref{eq:L_prob}), is complex, and thus it remains unclear if a robust model can be developed.

Machine learning offers an efficient approach to such cumbersome modelling
\citep{Moriwaki23}. 
Several studies have already explored the use of machine learning techniques to model the relationship between DM halos and galaxies \citep[e.g.,][]{Kamdar16,Jo19,Jespersen22}. While deterministic regression models are often used in these studies, it may be worth considering probabilistic models especially for the properties related to star formation activity, which is inherently stochastic.
Generative models \citep{Foster19} aim to learn the underlying probability distribution of the data to sample new data and are therefore well suited for this purpose.
This approach has recently been employed to populate dark matter halos with galaxies \citep[e.g.,][]{Lovell23, Nguyen24}. The typical strategy in these studies is a two-step process: one model first samples the number of galaxies within a given halo, and subsequently, a second model generates that number of galaxies.

In this study, we propose a generative model based on the Transformer architecture \citep{Vaswani17} to mainly model star formation rates (SFRs) of galaxies.
The Transformer, which relies on attention mechanisms, has gained significant attention in recent years due to its remarkable success across a wide range of natural language processing tasks \citep[e.g.,][]{Devlin18, Brown20}.
The attention mechanism enables the model to capture dependencies among all input elements within a single layer, thereby allowing efficient modelling of long-range dependencies. We develop a model that generates galaxies auto-regressively, simultaneously determining both their number and their individual properties within a single process.

In the following, we describe the model architecture and training dataset in \cref{sec:methods}. After presenting general results in \cref{sec:results}, we demonstrate that our models can generate larger map than the training data or lightcones when applied to DM-only simulations in \cref{sec:pinocchio}.
Finally, we discuss future prospects in \cref{sec:discussion}. 
Throughout this paper, we adopt a flat $\Lambda$ Cold Dark Matter (CDM) universe and use the following cosmological parameters: $\Omega_{\rm m} = 0.3089$, $\Omega_{\Lambda} = 0.6911$, $h = 0.6774$, $\sigma_8 = 0.8159$, and $n_\mathrm{s} = 0.9667$. 

\section{Methods}
\label{sec:methods}

\subsection{Model}

\label{sec:model}

\begin{figure}
    \centering
    \includegraphics[width=8cm]{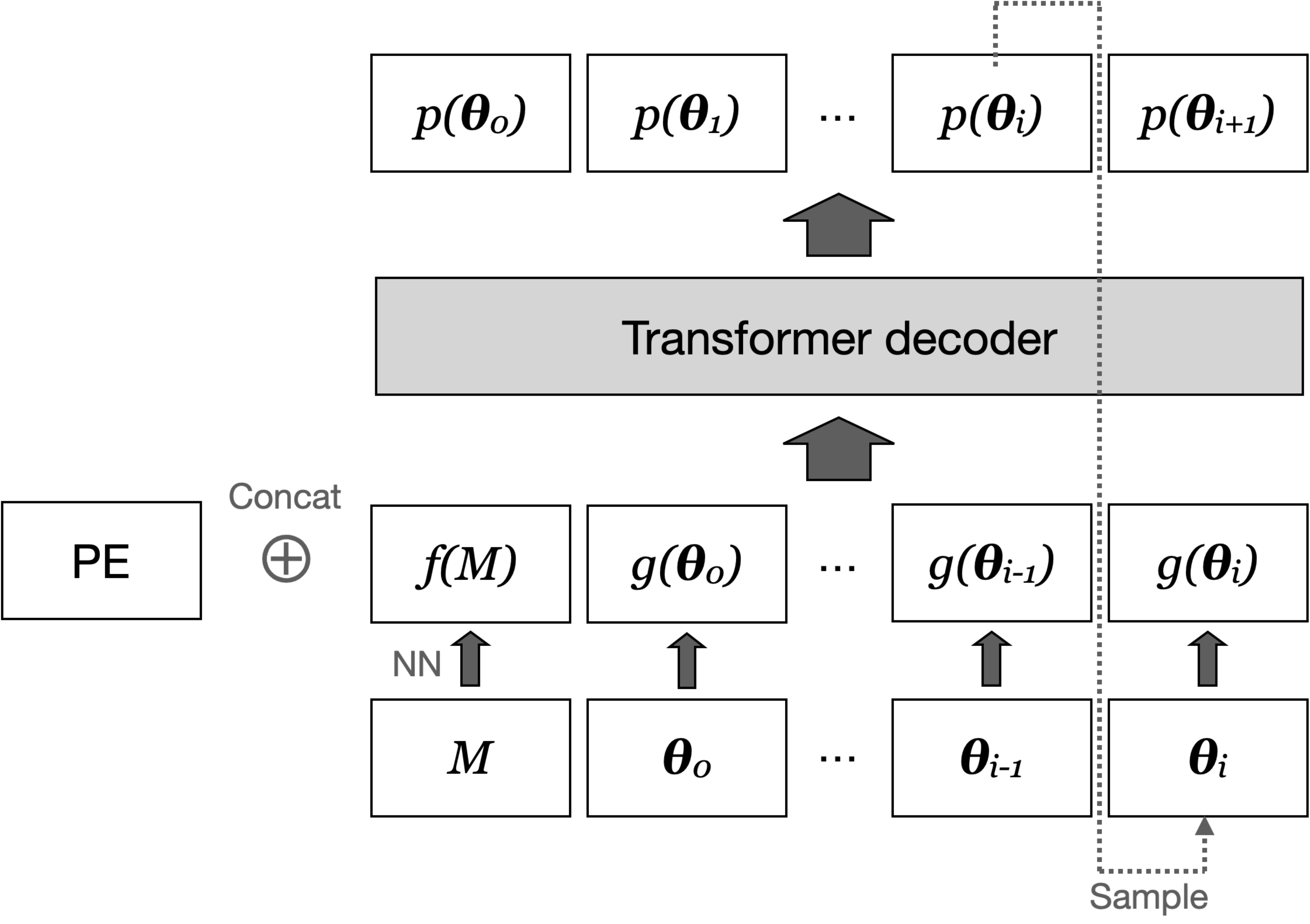}
    \caption{Schematic picture of the model architecture. The model takes the halo mass $M$ and the sequence of galaxy properties $\bm{\theta}$ as inputs. These inputs are embedded into a high-dimensional latent space using neural networks $f$ and $g$. The resulting sequence is passed through a Transformer decoder, which outputs a sequence of predicted probability density functions $(p(\bm{\theta}_0), \dots, p(\bm{\theta}_{i}))$. Positional encodings (PE) are concatenated to the input sequence to enable the model to learn the order of the galaxies. 
    For the prediction of $p(\bm{\theta}_{i})$, {\it future} elements, $\bm{\theta}_{i}$, $\bm{\theta}_{i+1}$, ..., are masked out and not used even if they are present in the input sequence.
    In the Transformer decoder, the self-attention mechanism allows all the values from $M$ to $\bm{\theta}_{i-1}$ to effectively contribute to the prediction of $p(\bm{\theta}_{i})$ within a few to several layers.
    When sampling new galaxies, the model generates them one by one, starting from the halo mass; at each step, the predicted probability distribution is used to sample the next galaxy property, as indicated by the dotted arrow. }
    \label{fig:architecture}
\end{figure}

We consider a set of galaxy properties, denoted by $\bm{\theta}$, 
and develop a generative model that produces a sequence $\{\bm{\theta}_i\}$ for a given halo mass $M$, where $\bm{\theta}$ includes SFR, distance to the halo centre, and radial and tangential velocities relative to the halo. 
We assume that the first galaxy in the sequence is the central galaxy and that subsequent satellite galaxies are sorted in descending order of SFR such that ${\rm SFR}_i \geq {\rm SFR}_{i+1}$ for $i = 2, 3, \cdots$. Note that the SFR of the central galaxy $\mathrm{SFR}_1$ is not always the highest among all galaxies in the halo.

To model the dependencies between member galaxies of a halo, we employ a Transformer-based architecture \citep{Vaswani17}. 
\Cref{fig:architecture} illustrates the architecture of our model. 
We first embed the halo mass and galaxy properties into a high-dimensional space using two separate neural networks, $f$ and $g$, respectively.
The resulting sequence, $(f(M), g(\bm{\theta}_0), ..., g(\bm{\theta}_{i-1}))$, is concatenated with learnable positional encodings and then passed into a Transformer decoder, which outputs a sequence of predicted probability density functions $(p(\bm{\theta}_0), \dots, p(\bm{\theta}_i))$. The positional encodings are used to let the model know the order of the galaxy properties. 
In the Transformer decoder, the self-attention mechanism allows all values from $M$ to $\bm{\theta}_{i-1}$ to effectively contribute to predicting $p(\bm{\theta}_i)$, with only a few to several attention layers.

We adopt two-layer fully connected feedforward networks with a hidden dimension of 128 and the LeakyReLU activation function for the embedding functions $f$ and $g$.
The Transformer decoder consists of a stack of self-attention layers, each comprising a multi-head attention and a feed-forward network.
We set the hidden dimension of the Transformer to 128, the number of attention heads to 8, and the number of self-attention layers to 4.

A common non-parametric approach to modelling probability distributions is to discretise the parameter space and predict the probability of each bin \citep[e.g.,][]{Ho21,Lima22,Stiskalek22,Rodrigues23}.
However, estimating a joint probability distribution $p(\bm{\theta})$ using binning becomes computationally intensive as the number of parameters increases.\footnote{
We have confirmed that the assumption is reasonable for the parameter sets considered in this paper, as described in \cref{app:joint_prob}.
We have also tested a normalizing flow \citep[][]{Rezende15, Kingma16} model, a class of generative models that are particularly well-suited for low-dimensional data and thus are expected to be appropriate for our task.
However, we have observed unstable behaviours to occur during its training, often with requiring a longer sampling time. Further details of our experiments are presented in \cref{app:nf}.
Prioritising simplicity and ease of use, we adopt a simple binning strategy in the present study.}
Therefore, we simply assume independence among the parameters and estimate each marginalized distribution individually. 
This allows us to focus on the core functionality of the model without being entangled in the complexities of the joint distributions. 
We set the number of bins to 100 and apply a softmax activation at the output of the Transformer decoder.

We train the model with a batch size of $B = 512$.
Each training batch consists of sequences of varying lengths, $\{(M^{(b)}, \bm{\theta}_1^{(b)}, \dots, \bm{\theta}_{l_b}^{(b)}, \bm{\theta}_{l_b+1}^{(b)})\}_{b=1}^B$, where $l_b$ is the number of member galaxies in the $b$-th batch,
$\bm{\theta}_{l_b}$ is the last element in the training data, and $\bm{\theta}_{l_b+1}^{(b)}$ is the end token that tells the model when to stop generating galaxies. 
We used a zero vector as an end token.  
We set the maximum length of the sequence to be 50. This number is somewhat arbitrary, but we find that most haloes host fewer galaxies, and even when more are present, their contribution to the observations is negligible.
The model is trained by minimizing the negative log-likelihood of the ground truth galaxy properties. Specifically, the loss function is defined as
\begin{align}
    \mathcal{L}_{\rm nll} = -\frac{ \sum_{b=1}^B \sum_{i=1}^{l_b+1} \log p(\bm{\theta}_i^{(b)})}{\sum_{b=1}^B (l_b+1)}.
\end{align}
We use the Adam optimizer \citep{Kingma14} and adopt a learning rate of $10^{-4}$ and a weight decay of $10^{-3}$. We also employ a cosine annealing learning rate scheduler \citep{Loshchilov16} with a minimum learning rate of $10^{-6}$ to ensure stable convergence. 
During training, we use a sampler that assigns higher sampling weights to underrepresented halo masses. The model is trained for 40 epochs. See \cref{app:hyperparameter} for the choice of hyperparameters.
The model is implemented in \texttt{PyTorch} \citep{pytorch}. The training for 40 epochs took about 20 minutes on a single NVIDIA RTX 6000 Ada Generation GPU. 

The galaxies are sampled from the predicted probability distribution one by one, starting from the halo mass, as:
\begin{equation}
\begin{aligned}
    \bm{\theta}^{\rm gen}_0 &\sim p(\bm{\theta}_0 | M), \\
    \bm{\theta}^{\rm gen}_1 &\sim p(\bm{\theta}_1 | M, \bm{\theta}^{\rm gen}_0),  \\
    \vdots 
\end{aligned}
\end{equation}
More specifically, we first select a single parameter bin according to the predicted categorical distribution, and then sample a value uniformly within the chosen bin.
The generation is terminated when the primary parameter (SFR in our case) takes a value smaller than a specified threshold. In this way, the number of galaxies assigned to each halo is automatically determined. To prevent unrealistic parameter combinations, the end-token probabilities of the other parameters are manually set to zero. We adopt the minimum SFR value in the training data of $10^{-3}~\rm M_\odot/yr$ as the threshold.

We observe that the model assigns non-zero probability even to values far outside the training distribution. This is likely due to the use of the softmax function, which rarely assigns exact zeros.
Although the probabilities of such unrealistic values are generally smaller than $10^{-4}$, they can still have a non-negligible impact on the observational statistics, especially when unrealistically large SFRs are sampled.
To mitigate this issue, we assign zero probability to SFR bins that are significantly greater than both the maximum values in the training data and the SFR of the previous satellite galaxy during sampling. We also treat the probabilities below $10^{-5}$ as zero.

\subsection{Dataset}
\label{sec:dataset}

We employ the cosmological hydrodynamic simulation IllustrisTNG \citep{Nelson19} as our training dataset. Among the simulations with different resolutions in the IllustrisTNG suite, we use TNG300-1 (hereafter, TNG for simplicity)\footnote{We do not use the DM-only counterpart to obtain DM halo properties because the haloes are not always identified nor matched properly. This leads to inconsistencies, especially when determining velocities and distances from the halo centre. We will examine the results of applying the trained model to DM-only simulations in the later sections.}, which has a box size of $(302.6 \, \mathrm{Mpc})^3$ and a mass resolution of $2.8 \times 10^8 \, \rm M_\odot $ for dark matter and $5.9 \times 10^7 \, \rm M_\odot $ for gas.
The subgrid models in TNG are calibrated to reproduce several observational results, including the cosmic SFR history, stellar mass function and stellar-to-halo mass relation at $z = 0$ \citep{Pillepich18a}. The simulation results are consistent with the observed galaxy properties including the galaxy clustering at $z < 2$ \citep{Springel18}, and some more comparisons with observations are made in, e.g., \citet{Springel18,Nelson18a,Naiman18,Marinacci18}.
In TNG, haloes are identified with a friends-of-friends (FoF) group finder, and within each halo, substructures, i.e., galaxies, are further identified using the \textsc{SubFind} algorithm \citep{Springel01}.

We consider four galaxy properties: SFR, distance to the halo centre $d$, and radial and tangential velocities relative to the halo $v_r$ and $v_t$. We denote the galaxy property data vector as $\bm{\theta} = ({\rm SFR}, d, v_r, v_\theta)$.
Many emission lines are tightly correlated with SFR \citep[e.g.,][]{Kennicutt98}, making modelling the SFR a crucial first step in modelling the luminosities of many lines.
The inclusion of the spatial distribution of galaxies in a halo and their velocities is important in LIM observations where either the angular or spectral resolution is high. 
The radial velocity is defined such that the velocity towards the halo centre is defined as positive, i.e., when a galaxy undergoes infall, the radial velocity is positive.
For the central galaxies, we set the distance to the halo centre and the radial velocity to be zero.
In TNG, the halo centre is defined by the position of the gravitational potential minimum. 

We use haloes with $M > 10^{11} ~\rm M_\odot$ and galaxies with SFR $> 10^{-3} ~\rm M_\odot / yr$. This results in a total of 330,287 haloes and 845,631 galaxies at $z = 2$.
For the satellite galaxies, we sort them in descending order of SFR. This is useful when one wants to generate only galaxies above a given SFR threshold.
For the test set, we reserved all haloes located within a subvolume of $(151.3~\rm Mpc)^3$ in TNG, corresponding to 12.5\% of the entire volume, which contains 38,746 haloes and 99,007 galaxies. From the remaining haloes, we further split the data into training and validation sets with a 9:1 ratio.

A DM-only run corresponding to TNG300-1 is available, named TNG300-1-Dark set (hereafter TNG-Dark). 
Matched catalogues of subhalos are also provided, but
we find that the matching is not always reliable; some galaxies are linked to haloes that are not their actual hosts, resulting in unrealistically large distances and relative velocities.\footnote{Among the 2,329,308 galaxies with SFR $ > 10^{-3}~\rm M_\odot /yr$ in the hydrodynamical run, we find that 28,752 (1.2 \%) and 301,076 (12.9 \%) galaxies lack a suitable counterpart when using the SubLink \citep{Rodriguez-Gomez15} and LHaloTree \citep{Nelson15} catalogues, respectively. These fractions do not change significantly even when only galaxies with higher SFR are considered (e.g., 2.5 \% and 13.1 \% for galaxies with threshold SFR $1~\rm M_\odot/yr$). Note that these numbers include galaxies residing in low-mass haloes. When restricting the sample to galaxies with subhalo masses (rather than host halo masses) greater than $10^{10}~\rm M_\odot$, we obtain very small fractions of $5\times 10^{-3}$ \% (SubLink) and 1.8 \% (LHaloTree), respectively. For future, more detailed analyses, it will be essential to ensure complete subhalo-host halo associations. %Since our model requires only subhalo-host halo assocciations rather than subhalo-subhalo matching, this can be . For more detailed analysis in future, matching 
} Although it is, in principle, possible to reassign the miss-matched galaxies to the correct haloes one by one, such a task is technically involved and is beyond the scope of the present study. We therefore use haloes in the hydrodynamics simulation for training.

We normalize the halo and galaxy properties before feeding them into the model. Specifically, we apply the following transformations:
\begin{align}
    \tilde{\theta} \to \frac{\tilde{\theta} - a_\theta}{ b_\theta - a_\theta} \quad (\theta = M,\ {\rm SFR},\ d,\ v_r,\ v_\theta),
\end{align}
where 
\begin{align}
    &\tilde{M} = \log_{10} M, 
    && a_M = 11, 
    && b_M = 15,\\
    &\tilde{\rm SFR} = \log_{10} {\rm SFR}, 
    && a_{\rm SFR} = -3.2, 
    && b_{\rm SFR} = 2.9,\\
    &\tilde{d} = \log_{10} d, 
    && a_d = -3,
    && b_d = 1,\\
    &\tilde{v}_r = {\rm sgn} (v_r) \log_{10}( |v_r| + 1) , 
    && a_{v_r} = -4 ,
    && b_{v_r} = 4 ,\\
    &\tilde{v}_\theta = \log_{10} |v_\theta| ,
    && a_{v_\theta} = -2 ,
    && b_{v_\theta} = 4,
\end{align}
with $M$, SFR, $d$, $v_r$, and $v_\theta$ given in units of ${\rm M_\odot}$, ${\rm M_\odot / yr}$, ${\rm Mpc}/h$, km/s, and km/s, respectively.
Note that the radial velocity, $v_r$, has different physical representation when positive (infall) and negative (receding from the halo centre). We therefore adopt a normalisation scheme that retains the sign information.
The normalisation parameters $a$ and $b$ are chosen so that the normalised values of the training data fall within the range of $(0, 1)$.

\section{Results}
\label{sec:results}

\subsection{Comparison with TNG galaxies}

\begin{figure}
    \centering
        \includegraphics[width=8.5cm]{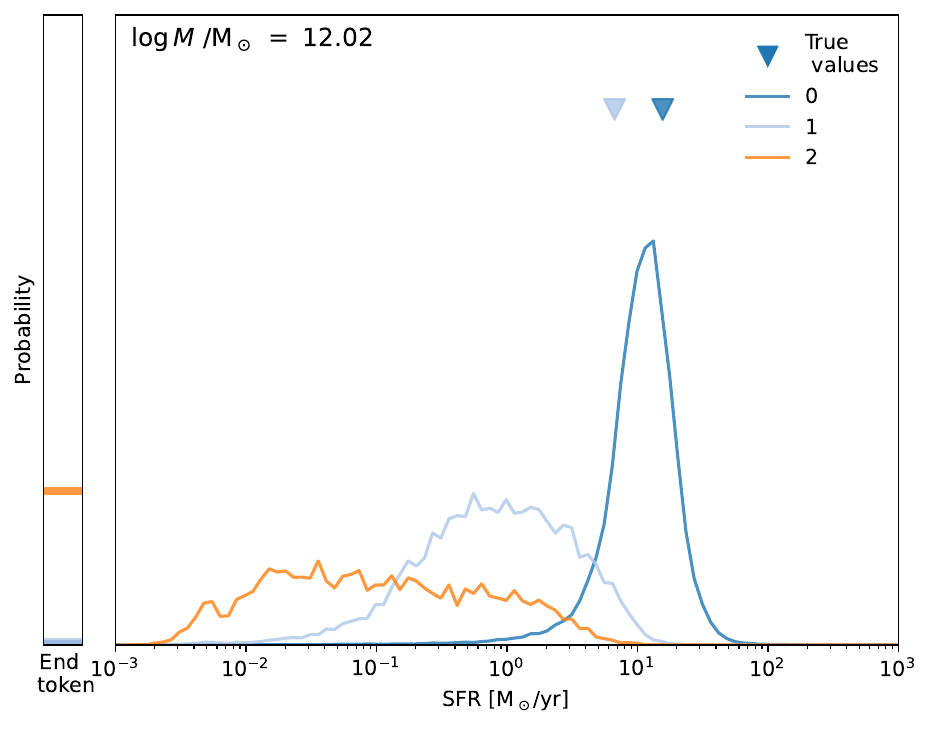}
        \includegraphics[width=8.5cm]{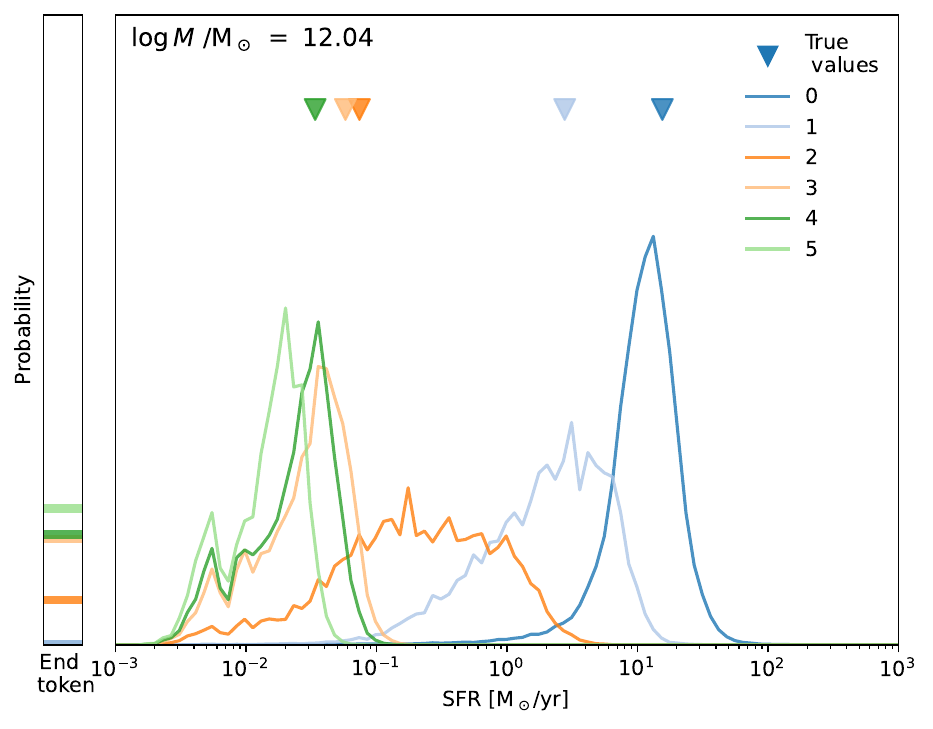}
    \caption{
    Two examples of the predicted probability distributions of SFR for test haloes with $M \sim 10^{12} ~\rm M_\odot$. The inverted triangle indicates the true values, which are used to predict the probability distributions. The narrow panels on the left show the probabilities of sampling the values below $10^{-3}~\rm M_\odot/yr$, which are considered as end token, on a linear scale spanning from 0 at the bottom to 1 at the top. The numbers in the labels indicate order within haloes: 0 for the central, 1 for the first satellite, and so on. The top and bottom examples contain two and five galaxies, and the predicted probability densities are shown up to the third and sixth galaxies, respectively.}
    \label{fig:prob_true}
\end{figure}

The top and bottom panels of \cref{fig:prob_true} show two examples of the predicted probability distributions of SFR for test TNG haloes with $M \sim 10^{12} ~\rm M_\odot$. 
The inverted triangles show the true values.
Since the input halo mass is almost the same, the model returns almost the same probability distributions for the central galaxies (blue; labelled as 0).
However, because the sampled properties are slightly different, the probability distributions of the first satellite galaxies (light blue; labelled as 1) are different. 
We further show the probability distributions for the other galaxy properties as well as those obtained auto-regressively using the generated galaxies in \cref{app:prob}.

The fact that the probability distributions have non-zero width indicates that the model is not over-fitting specific samples in the training data, and that it behaves non-deterministically.
This probabilistic nature ensures that the model can always generate different realisations of galaxies even when using the same halo catalogue.

\begin{figure}
    \centering
    \includegraphics[width=8cm]{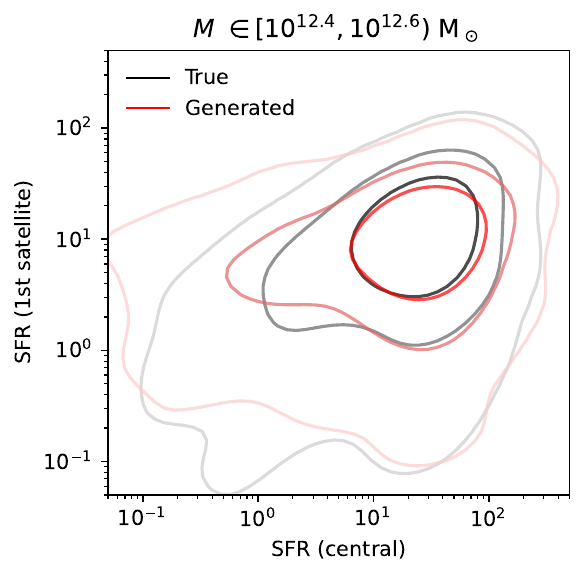}
    \caption{Two-dimensional distributions of central and first satellite SFRs for the TNG (black) and generated (red) catalogues. Haloes within $M \in [10^{12.4}, 10^{12.6}) ~\rm M_\odot$ are considered. The contours represent 0.5, 1, and 2-$\sigma$ levels.}
    \label{fig:contour}
\end{figure}

Next, we investigate the correlation between the SFRs for central and satellite galaxies. 
We find that this correlation is not very clear at $M \lesssim 10^{12}~\rm M_\odot$ but becomes more evident at higher halo masses in the simulation. 
\Cref{fig:contour} shows the distributions of haloes with halo masses around $M\sim 10^{12.5}~\rm M_\odot$, where the horizontal and vertical axes correspond to the SFRs of central and first satellite galaxies, respectively.  
The black and red contours represent the 0.5, 1, and 2-$\sigma$ levels for the test TNG galaxies and those generated from the same test haloes. We find a modest correlation in both TNG and our generated catalogues. 
To compare the two distributions, we compute the 1-Wasserstein distance \citep{Peyre18}, a measure of the discrepancy between probability distributions, using the python library {\sc pot} \citep{Flamary21_POT,Flamary24_POT}\footnote{https://github.com/PythonOT/POT}. As a reference, we also compute the distance between the true samples and a control set obtained by randomly shuffling the first satellite SFR of the true samples to remove the correlation between the central and satellite SFRs. We find that the distances are 0.024 for the generated samples and 0.036 for the shuffled samples. 
The smaller distance for the generated samples indicates that our model successfully captures a key aspect of the galaxy correlations encoded in the simulation.
Additional comparisons between the true and generated galaxies are presented by visualizing the joint parameter distributions in \cref{app:joint_prob} and by conducting a coverage-based test in \cref{app:tarp}.

\subsection{Generating mock data from TNG haloes}

In this section, we present mock data generated with our model. As we consider SFR instead of line luminosities, we generate an SFR density map, where each voxel contains the total SFR divided by its volume, as a close proxy of a line intensity map.
As discussed in the previous section, our probabilistic generative approach allows the model to produce different realisations of galaxies even when using the same halo catalogue.
It is therefore of interest to compare the newly generated mock data with the original TNG haloes, not only on test data but also on training data. 

In the following, we show the results generated from TNG-Dark haloes alongside those generated from TNG haloes. 
Baryonic effects are known to reduce halo masses. For example, \citet{Sawala13} demonstrated that at $M\sim 10^{12}~\rm M_\odot$, haloes in hydrodynamics simulations have masses a few percent smaller than the DM-only counterparts. 
To account for this effect, we adopt a simple prescription:\footnote{More detailed corrections, such as those depending on halo properties, can be made using the matching catalogue.} when applying our model to DM-only simulations, we rescale the original halo mass, $M_{\rm DMO}$, by a factor $\eta$, such that $M = \eta M_{\rm DMO}$. We adopt $\eta = 0.9$ so that we find reasonably consistent halo mass functions.

To populate galaxies, we place each galaxy at radius $d$ along a random direction from the centre.
We find it takes about $20$ seconds to generate all galaxies in TNG with $\rm SFR > 10^{-3}~M_\odot /yr$ for haloes with $M > 10^{11}~\rm M_\odot$ at $z = 2$ with a batch size of 512. 

\begin{figure*}
    \centering
    \includegraphics[width=16cm]{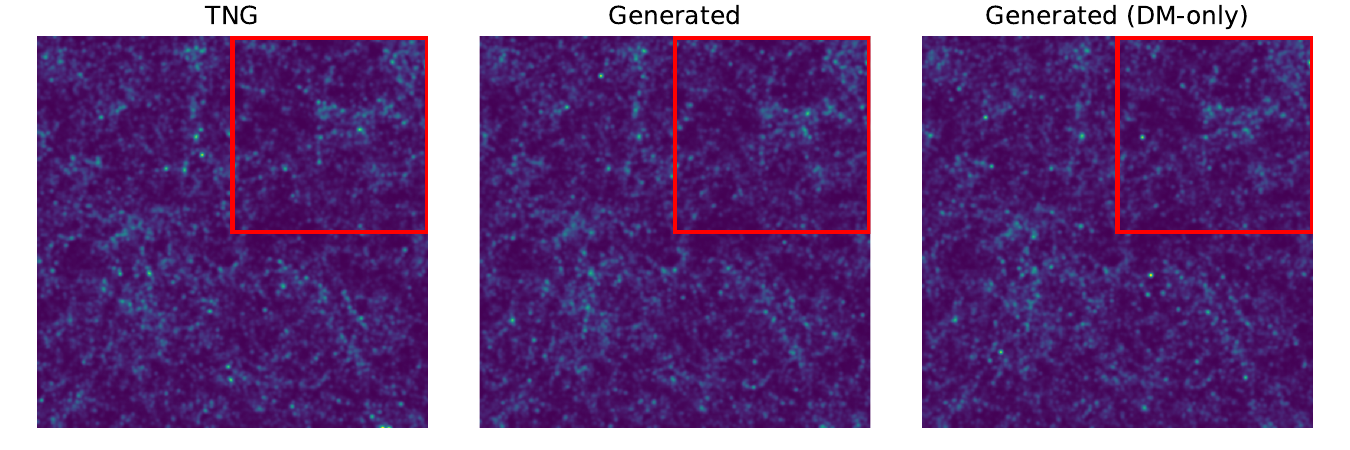}
    \caption{SFR maps constructed from TNG galaxies (left) and galaxies generated from haloes in TNG (middle) and TNG-Dark (right).
    Each map has a pixel resolution of 0.59 Mpc.
    We do the projection over the half of the simulation volume ($302.6 \times 302.6 \times 151.3$ Mpc) that contains the test region. 
    The region reserved as test data is indicated by the red box.
    The images are smoothed with a Gaussian kernel for visual clarity. }
    \label{fig:intensity_maps}
\end{figure*}

We divide the volume into $512^3$ voxels, each with a size of $(0.59 \ \rm Mpc)^3$, which corresponds to an angular resolution of 23 arcsec at $z = 2$.
\Cref{fig:intensity_maps} shows the SFR maps of the TNG galaxies (left), and the galaxies generated from haloes in TNG (middle) and TNG-Dark (right).
To ensure consistency, only haloes with $M > 10^{11} ~\rm M_\odot$ are used.
The region not used for training is indicated by the red boxes.
The generated mock data exhibits a similar large-scale distribution, but a different small-scale realisation even within the regions used for training. This again demonstrates the ability of our model to generate diverse realisations from the same halo distribution.

\begin{figure}
    \centering
    \includegraphics[width=8.5cm]{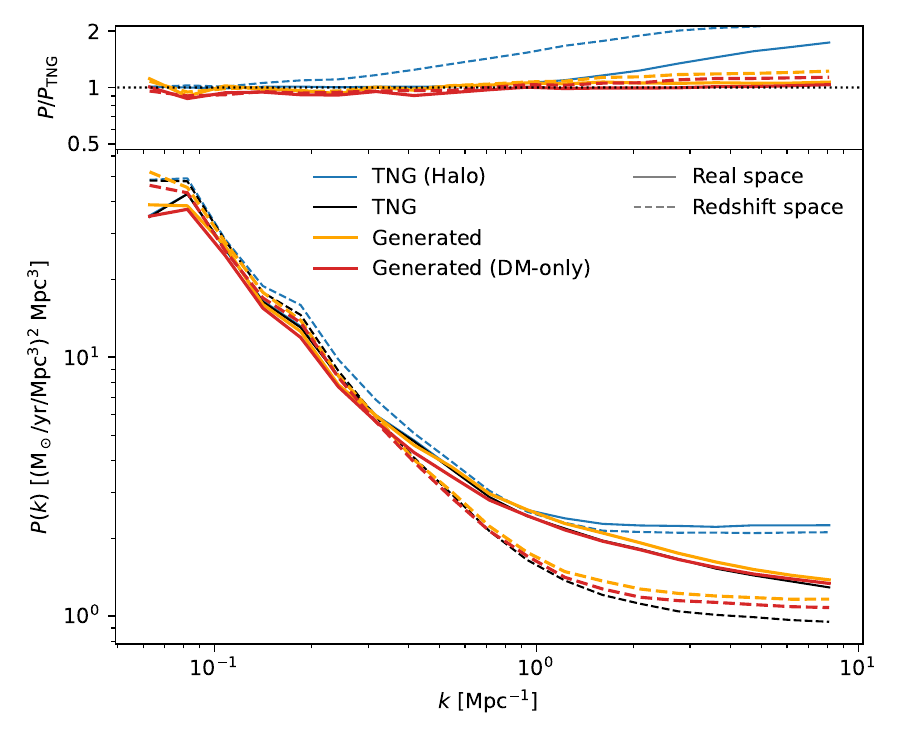}
    \caption{Power spectra of SFR maps of TNG haloes (light blue), TNG galaxies (black), and the mock galaxies generated from TNG haloes (orange) and TNG-Dark haloes (red) in the test region. The solid and dashed lines indicate the power spectra in real and redshift space, respectively. }
    \label{fig:power}
\end{figure}

\begin{figure}
    \centering
    \includegraphics[width=8.5cm]{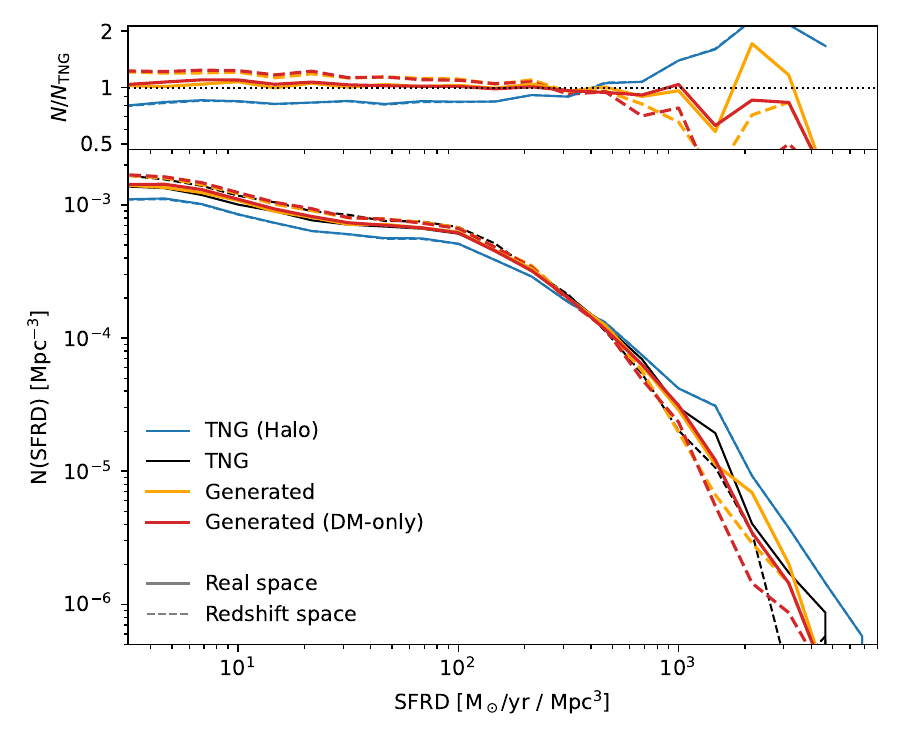}
    \caption{Voxel SFR density distributions for TNG haloes (light blue), TNG galaxies (black), and the galaxies generated from TNG (orange) and TNG-Dark (red) in the test region. The solid and dashed lines indicate the histograms of the SFR density in real and redshift space, respectively.}
    \label{fig:VID}
\end{figure}

\Cref{fig:power} compares the angle-averaged power spectrum of the generated mock data with that of the true TNG data (black). The orange and red lines present the generated results with TNG and TNG-Dark, respectively. The top panel of \cref{fig:power} shows the ratio of the generated power spectra to the true TNG power.
We find that the power spectrum is properly reproduced when galaxies are generated from TNG haloes, and also when using TNG-Dark haloes, provided that an appropriate halo-mass rescaling factor is applied.

In the context of LIM, a commonly used method for populating galaxies is to assign a single luminosity to the halo centre \citep{Yue15, Silva15, Moriwaki20, Bethermin22, Gkogkou23, Roy23, Mas-Ribas23}. 
As a simplified proxy for this approach, the light-blue line in \cref{fig:power} shows the power spectrum computed solely using the haloes of TNG, where the total SFRs of true member galaxies are assigned to the halo centres. Due to larger shot noise, the power spectrum computed with TNG haloes has a larger amplitude on small scales (large-$k$) than that from TNG galaxies.
In contrast, the generated mock data have power spectra similar to those of the TNG galaxies even on the smallest scales, indicating that the distribution of the galaxies within haloes is well captured.

Since our model can predict the velocities of galaxies, we can incorporate redshift space distortion (RSD) effects.
To do so, we compute the velocities of generated galaxies by adding the sampled velocities $v_r$ (directed toward the halo centre) and $v_t$ (with a randomly chosen direction perpendicular to the radial direction) to the halo velocity. 
In \cref{fig:power}, the dashed lines show the power spectrum with RSD. The generated data reproduce the characteristic RSD effect, where the power increases on large scales and decreases on small scales compared to the real-space power spectrum. Once again, the generated mock data exhibit a similar power spectra to those of TNG, indicating that the velocities of the member galaxies are well captured by our model.

\Cref{fig:VID} shows the voxel SFR density distribution --- a proxy for the voxel intensity distribution --- for TNG haloes (light blue), TNG galaxies (black), and the generated mock catalogues from TNG (orange) and TNG-Dark (red) in real (solid) and redshift space (dashed). The histogram of the generated mock data is consistent with that of the TNG haloes in both cases.

\begin{figure*}
    \centering
    \includegraphics[width=16cm]{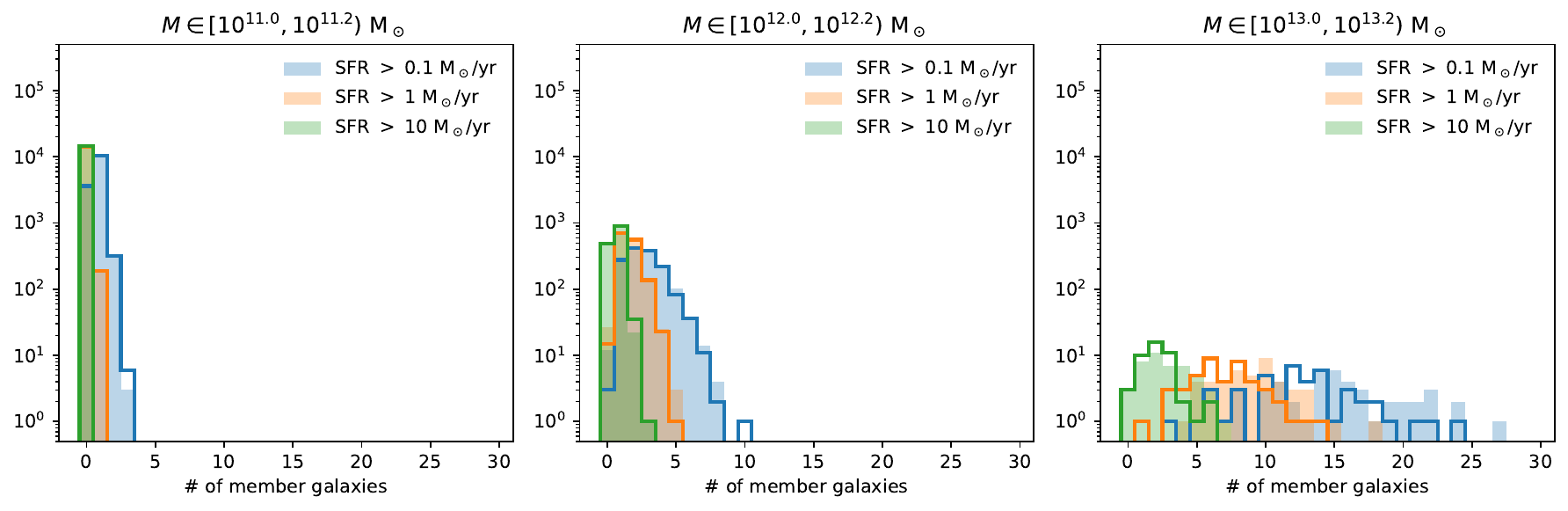}
    \caption{Histograms of the number of galaxies per halo with SFRs above 0.1 (blue), 1 (orange), and 10 $\rm M_\odot /yr$ (green) generated for haloes in mass bins around $M \sim 10^{11}$ (left), $10^{12}$ (middle), and $10^{13}~\rm M_\odot$ (right). The filled and line histograms represent the TNG and the generated galaxies, respectively, in test haloes. }
    \label{fig:hist_length}
\end{figure*}

Our model can also be used for spectroscopic cosmological surveys targeting emission-line galaxies \citep[e.g.,][]{Takada14, DESI16, Euclid25, LSST09, Roman15}.
In such cases, the number of galaxies above a certain threshold is crucial. \Cref{fig:hist_length} shows the histogram of the number of galaxies per halo with $\rm SFR >  0.1$ (blue), 1 (orange), and 10 (green) $\rm M_\odot /yr$, in three different mass bins. The filled and outline histograms represent the test TNG data and the generated galaxies. 
We find that the two distributions are consistent with each other.

\section{Application to DM-only simulations}
\label{sec:pinocchio}
\subsection{Generating mock data with larger volume}

\begin{figure*}
    \centering
    \includegraphics[width=15cm]{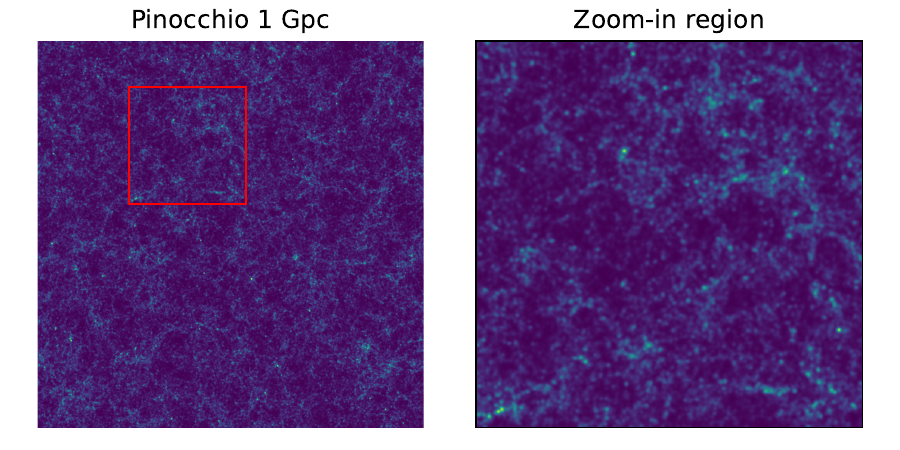}
    \caption{SFR maps constructed from {\sc pinocchio} haloes, showing the full 1 Gpc box (left) and a zoom-in view of a region with a side length of 302.6 Mpc (right). Each map has a pixel resolution of 0.59 Mpc and shows a projection over 151.3 Mpc along the line of sight.}
    \label{fig:intensity_maps_pinocchio}
\end{figure*}

\begin{figure}
    \centering
    \includegraphics[width=8.5cm]{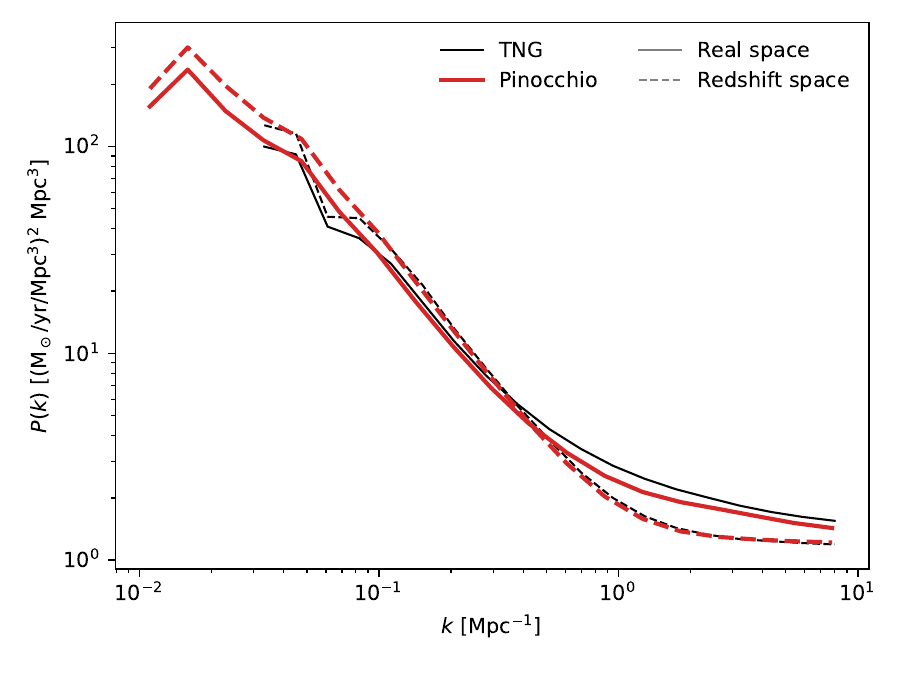}
    \caption{Power spectra of SFR density maps of TNG galaxies (black) and galaxies generated for {\sc pinocchio} haloes (red). The solid and dashed lines indicate the power spectra in real and redshift space, respectively.}
    \label{fig:power_pinocchio}
\end{figure}

To demonstrate the ability of our model to generate mock data with a larger volume, we apply it to a halo catalogue generated with the fast DM simulation code {\sc pinocchio} \citep{Monaco13}.
{\sc pinocchio} is an algorithm based on Lagrangian perturbation theory. 
We use a catalogue created with a particle mass of $6.7\times10^{9}\ {\rm M_\odot}$
with the mass rescaling factor $\eta = 0.98$.
The left panel of \cref{fig:intensity_maps_pinocchio} shows the SFR density map of the generated data for a box with 1 Gpc on a side, while the right panel shows a zoom-in view of a subregion with a side length of 302.6 Mpc.
\Cref{fig:power_pinocchio} further compares the power spectrum of the TNG galaxies (black) with that of the generated map based on the {\sc pinocchio} haloes (red). 
The {\sc pinocchio} power spectrum is consistent with the TNG galaxies on small scales, and smoothly extends to larger scales.

The trained model can be applied to much larger volumes, but care must be taken for dealing with massive haloes: machine-learning models are typically poor at extrapolation, particularly when applied to regimes that lie far outside the range of the training data.
The maximum halo mass in training data is $\log M$ = 14.2, and beyond this there is no guarantee that the model can generate galaxies properly. 
Although the most massive haloes tend not to significantly affect the power spectrum due to their rarity, they can affect pixel-by-pixel analyses. To handle such haloes, one can either adopt a separate model tailored to massive haloes or fine-tune the model using zoom-in simulations of massive haloes. 
When using zoom-in simulations for training, however, it is important to ensure that the model is not biased by unrepresentative populations. For instance, training on cluster regions alone would lead to generating only galaxies typical of such environments.
We note that the most massive halo in our 1 Gpc {\sc pinocchio} catalog has $\log M$ = 14.5, just slightly above the maximum value in our training data. Therefore, we do not expect any erratic or unphysical behavior in this particular application.

\subsection{Generating lightcone}
\label{sec:lightcone}

Creating lightcones is crucial in LIM, particularly for incorporating all possible emission lines observed at a given wavelength for wide survey areas.
Here, we demonstrate that this can be achieved by using models trained on simulation snapshots at different redshifts.

We train 15 models with galaxies at 15 different redshifts in the range $z = 0.5 \text{--} 6.0$. Among all the hyperparameters, we vary only the number of epochs from the value used for the $z = 2$ model. We test with 20, 40, 60, and 80 epochs and selected the models with the smallest validation errors.
We confirm that the model does not overfit, i.e., it generates diverse sets of galaxies for the same halo mass, and that the statistical properties including the power spectrum and voxel SFR density distribution are reasonably well reproduced across all redshifts. 

We apply these models to a {\sc pinocchio} lightcone halo catalogue, which was constructed by concatenating simulation volumes at different redshifts. We set the simulation box size to 600 Mpc$/h$ (= 857 Mpc), the aperture of the lightcone to 1.5 deg, and the minimum halo mass to $10^{11}~\rm M_\odot$. 
For each halo in the lightcone, galaxies were generated using the model trained at the redshift that is closest to that of the halo. 

\begin{figure*}
    \centering
    \includegraphics[width=17cm]{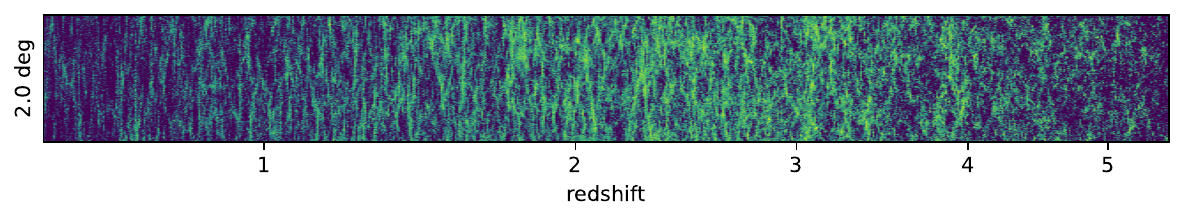}
    \caption{SFR density lightcone generated from a {\sc pinocchio} halo catalogue. The map shows a projection over 0.5 deg.}
    \label{fig:sfr_lightcone}
\end{figure*}

\begin{figure}
    \centering
    \includegraphics[width=8.5cm]{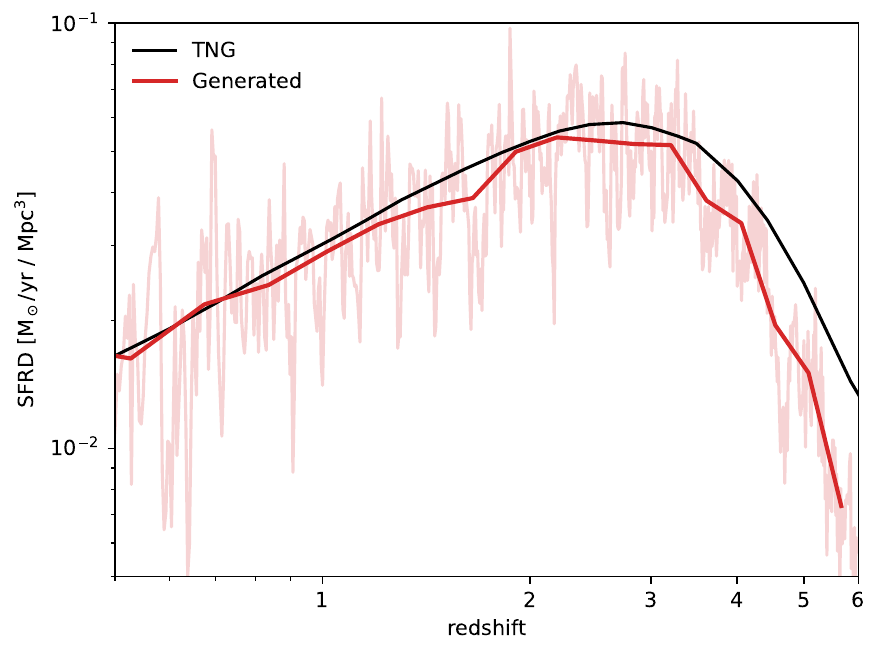}
    \caption{Redshift evolutions of the mean SFR densities of TNG300 (black) and generated lightcone (red). For the generated lightcone, the average over $\Delta \log z = 0.04$ is shown. The jagged faint red line corresponds to the result computed with a finer binning of $\Delta \log z = 0.001$.}
    \label{fig:z_sfrd}
\end{figure}

\Cref{fig:sfr_lightcone} shows an example of a generated lightcone, where the colour represents the voxel SFR density. 
We also show the mean SFR density evolution in \cref{fig:z_sfrd}, with the black and red lines representing the TNG simulation and the generated lightcone, respectively. The generated data closely follows the trend of the TNG, where the SFR density peaks at around $z = 2$. 
These results demonstrate the effectiveness of our redshift-dependent models in generating realistic lightcones for LIM studies.

\section{Discussion}
\label{sec:discussion}

We have developed a model that generates mock galaxy populations from dark matter halo catalogues. Compared to other methods with similar philosophy in the context of LIM \citep[e.g.,][]{Yue15, Moriwaki20, Bethermin22}, our approach aims to capture the complex correlations between galaxies by training a generative model on the hydrodynamics simulation data.
It should be noted, however, that the simulation is not necessarily the perfect representation of reality. Indeed, the outcomes of hydrodynamic simulations vary depending on the choice of subgrid models and resolution \citep[see, e.g.,][]{Vogelsberger20}.
To take into account a broader range of possible galaxy models and enhance the flexibility of our generative model, 
it may be an idea to use a suite of simulations such as CAMELS \citep{Villaescusa-Navarro23}, which explore a wide variety of subgrid models.
Such an approach has been taken in, e.g., \citet{Lovell23, Bourdin24, Kwon24, Nguyen24}.
The dependence on simulation modelling can be incorporated by conditioning the subgrid parameters, either jointly with the halo properties or through a separate context vector concatenated to the input sequence just as the positional encoding (see \cref{fig:architecture}).

Including halo properties beyond mass may further improve the model.
It has been shown that even for haloes with similar halo masses, galaxy properties further depend on their environment both in the local and high-redshift universe \citep[e.g.,][]{Peng10, Lemaux22}. 
Some HOD models for ELGs already incorporate such environmental effects \citep[e.g.,][]{Osato22, Hadzhiyska23}.
In addition, halo concentration plays a crucial role in shaping the SFRs and the abundance of satellite galaxies \citep{Bose19,Jun25b}, and therefore incorporating it as a secondary parameter may lead to more accurate modelling of galaxy-halo connections.

By generating additional galaxy properties such as metallicity, one can adopt more sophisticated models of line emission.
Moreover, generating other physical properties, such as stellar mass and black hole accretion rate, would be useful. This would enable the modelling of multiple galaxy populations, including luminous red galaxies and active galactic nuclei together. It can also be used for modelling continuum in LIM observations.
These physical properties are expected to be tightly correlated and therefore require careful joint modelling.

The mock data generated by the model developed in this study can be used for a variety of LIM applications, including observational forecasts and the development and validation of data analysis pipelines. 
It is also suitable for joint analyses across multiple surveys, including cross-correlations of LIM with galaxies \citep{Keenan22,Dunne25}, quasars \citep{Breysse19}, weak lensing maps \citep{Shirasaki21}, and CMB lensing maps \citep{Maniyar22,Fronenberg24}. 
While we demonstrated our model on halo catalogues with relatively small box sizes for illustrative purposes in \cref{sec:pinocchio}, it is also applicable to lightcones with much larger area such as those constructed and used in \citet{Hadzhiyska22,Smith22}.
Our new scheme for constructing halo-galaxy connection models using a probabilistic, autoregressive generative framework paves the way for realistic mock data generation for a broad range of large-scale structure studies.

\section*{Acknowledgements}

The authors acknowledge financial support from JSPS KAKENHI Grant Number JP23K03446, JP23K20035, JP24H00004 (KM), 
JP24H00215, JP25K17380, JP25H01513, JP25H00662 (KO), 
and JP24H00221 (NY).
This work was supported by MEXT as “Program for Promoting Research on the Supercomputer Fugaku'' (Structure and Evolution of the Universe Unraveled by Fusion of Simulation and AI; Grant Number JPMXP1020230406; Project ID: hp230204, hp240219, hp250226).

\section*{Data Availability}

The machine learning models developed in this study are available at \url{https://github.com/knmoriwaki/cosmoglint}.
The TNG simulation data used for training is publicly available.\footnote{\url{https://www.tng-project.org}}

%%%%%%%%%%%%%%%%%%%% REFERENCES %%%%%%%%%%%%%%%%%%
\bibliographystyle{mnras}
\bibliography{bibtex_library}

%%%%%%%%%%%%%%%%% APPENDICES %%%%%%%%%%%%%%%%%%%%%

\appendix

\section{Choice of Hyperparameters}
\label{app:hyperparameter}

Previous studies have shown that the hyperparameters of Transformer models only weakly affect the final results \citep[e.g.,][]{Kaplan20}. For instance, the model can generally achieve enough representational capacity as long as the model is deep enough and has large enough hidden dimensions. 
Based on these findings, we fixed the architecture of our model (e.g., the number of layers, attention heads, and hidden dimension) throughout this study.
We found that the model performed well even without extensive hyperparameter tuning. 
Such robustness of the model is particularly valuable when extending the model to datasets with different redshifts or galaxy properties.

For the training parameters, we first tuned the initial learning rate to ensure stable training.
Subsequently, with all other parameters fixed, we explored the optimal number of epochs and the choice of data sampler.
We used a custom sampler that assigns sampling weights inversely proportional to the frequency of samples in each halo mass bin. The weights are clamped to a minimum value ($w_{\rm min}$) to prevent excessively low weights for rare haloes. 
A larger $w_{\rm min}$ increases the sampling frequency of rare haloes, which can accelerate training but may also lead to fast overfitting.
We tested the model with $w_{\rm min} =$ 0.01, 0.02, and 0.05. 
We found that the values below 0.01 led to rapid overfitting, while values larger than 0.05 required longer time (more than an hour) to reach the minimum validation loss.
The optimal number of epochs for each $w_{\rm min}$ was approximately 20, 40, and 100. Although the final validation loss values were almost identical for all settings, the results for $w_{\rm min}=0.02$ and 0.05 were slightly lower than for $w_{\rm min} = 0.01$.
The model with $w_{\rm min} = 0.02$ with 40 epochs was adopted for the main results in this study.

\section{Additional evaluation}
\label{app:additional_evaluation}

\subsection{Predicted probability densities}
\label{app:prob}

\begin{figure*}
\centering
\begin{minipage}{0.33\textwidth}
    \centering
    \includegraphics[width=\linewidth]{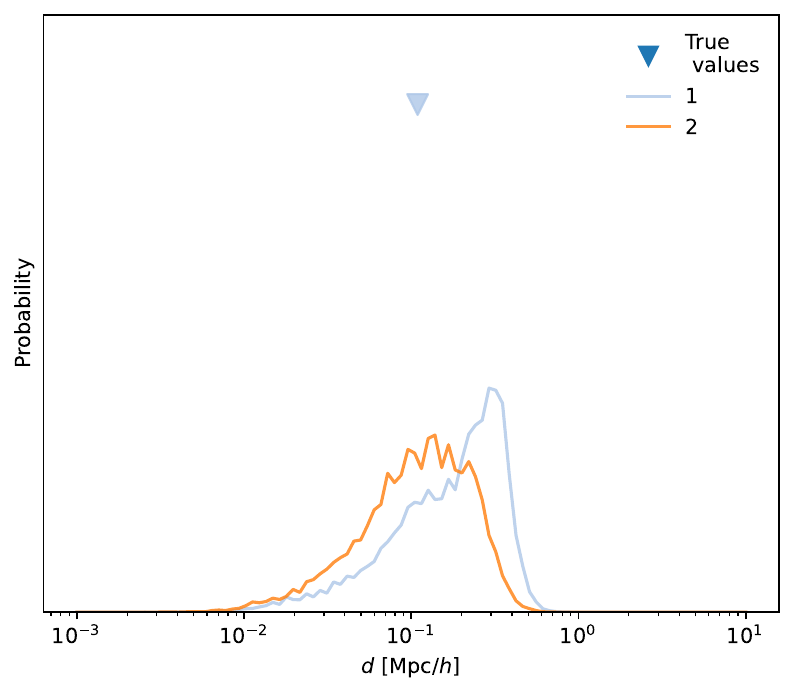}
\end{minipage}
\hfill
\begin{minipage}{0.33\textwidth}
    \centering
    \includegraphics[width=\linewidth]{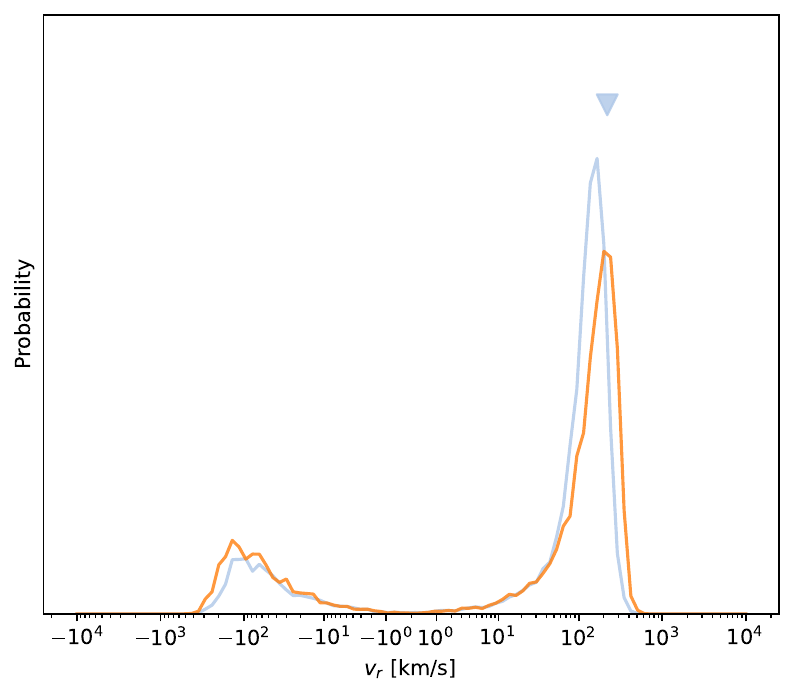}
\end{minipage}
\hfill
\begin{minipage}{0.33\textwidth}
    \centering
    \includegraphics[width=\linewidth]{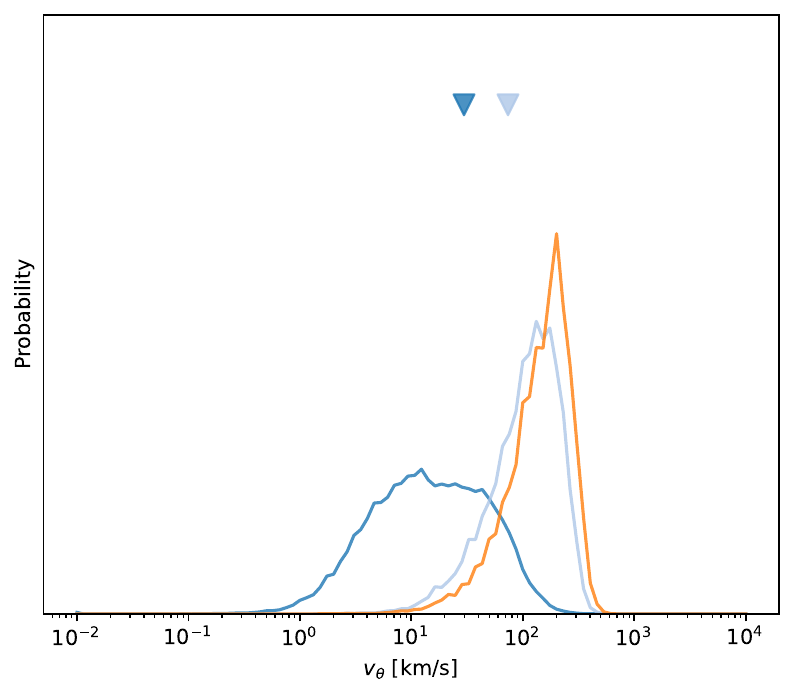}
\end{minipage}

\begin{minipage}{0.33\textwidth}
    \centering
    \includegraphics[width=\linewidth]{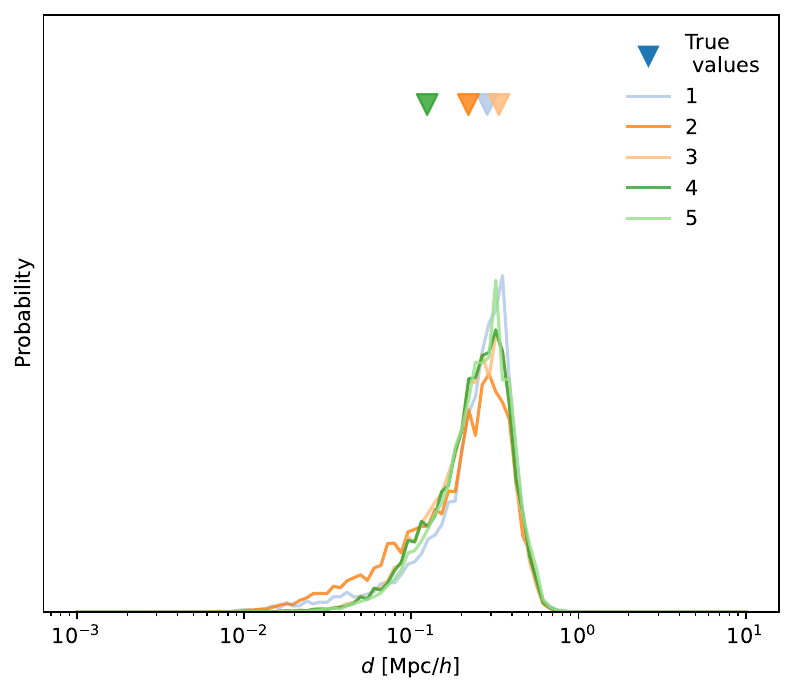}
\end{minipage}
\hfill
\begin{minipage}{0.33\textwidth}
    \centering
    \includegraphics[width=\linewidth]{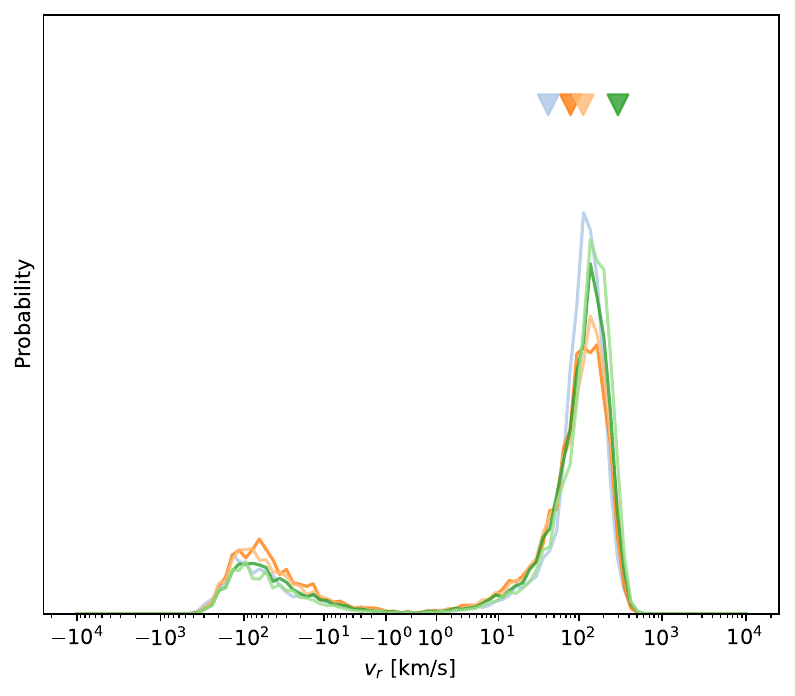}
\end{minipage}
\hfill
\begin{minipage}{0.33\textwidth}
    \centering
    \includegraphics[width=\linewidth]{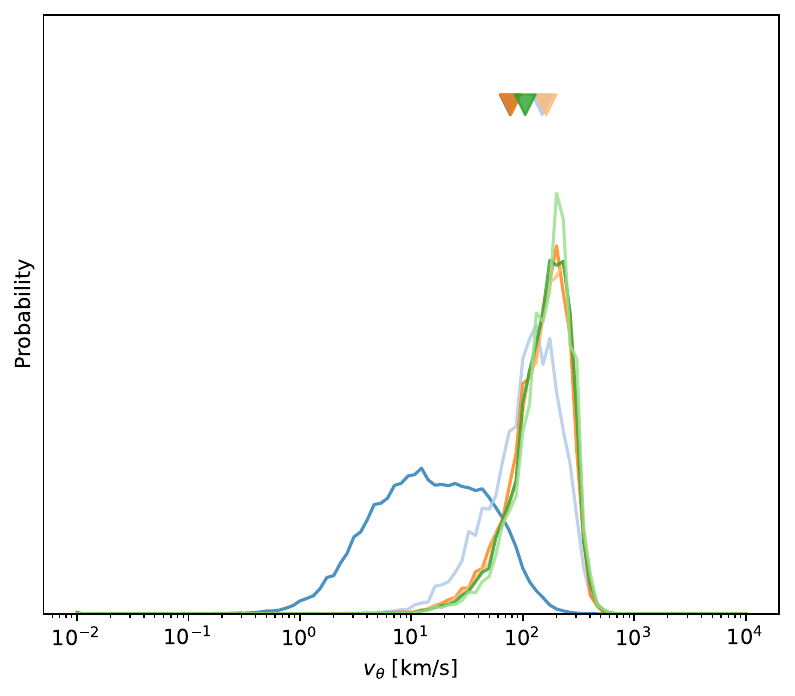}
\end{minipage}
\caption{
The probability distributions of the distance to the halo centre (left), radial velocity (middle), and tangential velocity (right) for the haloes shown in \cref{fig:prob_true}.
}
\label{fig:prob_true_d_vr_vt}
\end{figure*}

\Cref{fig:prob_true_d_vr_vt} shows the probability distributions of the distance to the halo centre (left), radial velocity (middle), and tangential velocity (right) for the haloes shown in \cref{fig:prob_true}.
For the distance and radial velocity, the values for the central galaxies (blue; labelled as 0) are fixed to zero, and the corresponding probabilities are therefore not shown.
Although the differences are less significant than in the case of SFR, we still observe halo-to-halo variations in these properties.

\begin{figure*}
\centering
\begin{minipage}{0.255\textwidth}
    \centering
    \includegraphics[width=\linewidth]{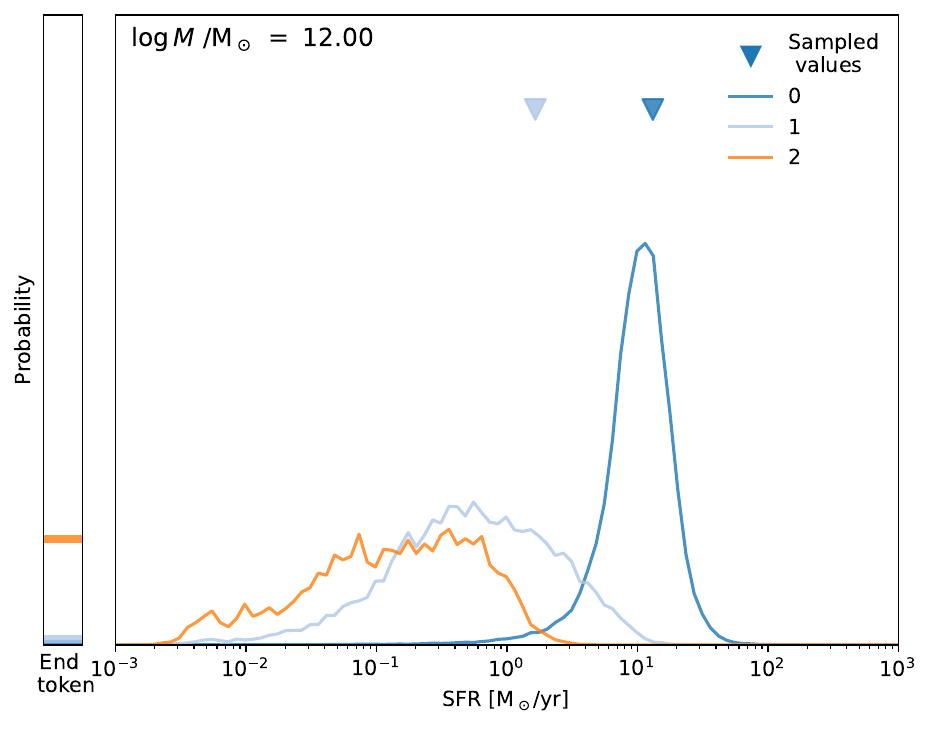}
\end{minipage}
\hfill
\begin{minipage}{0.23\textwidth}
    \centering
    \includegraphics[width=\linewidth]{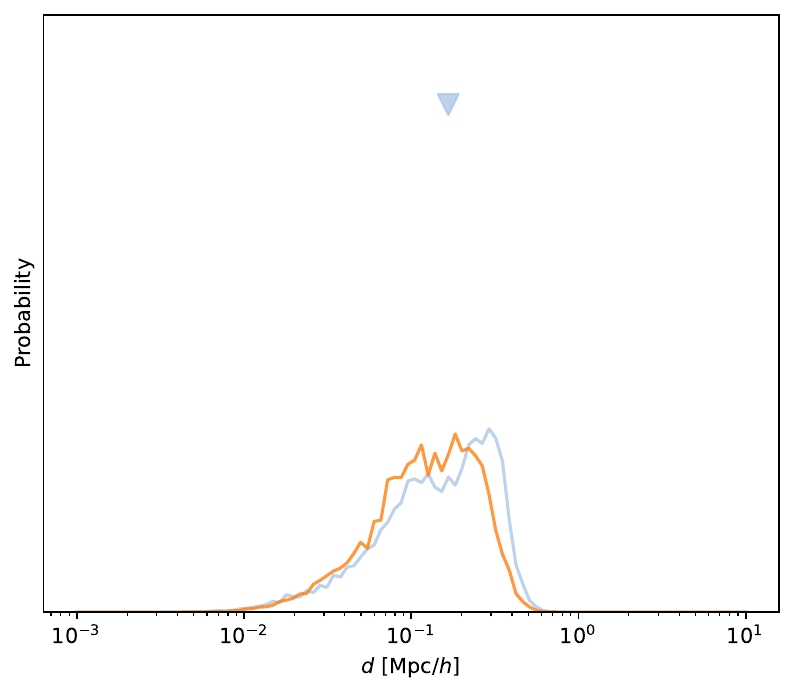}
\end{minipage}
\hfill
\begin{minipage}{0.23\textwidth}
    \centering
    \includegraphics[width=\linewidth]{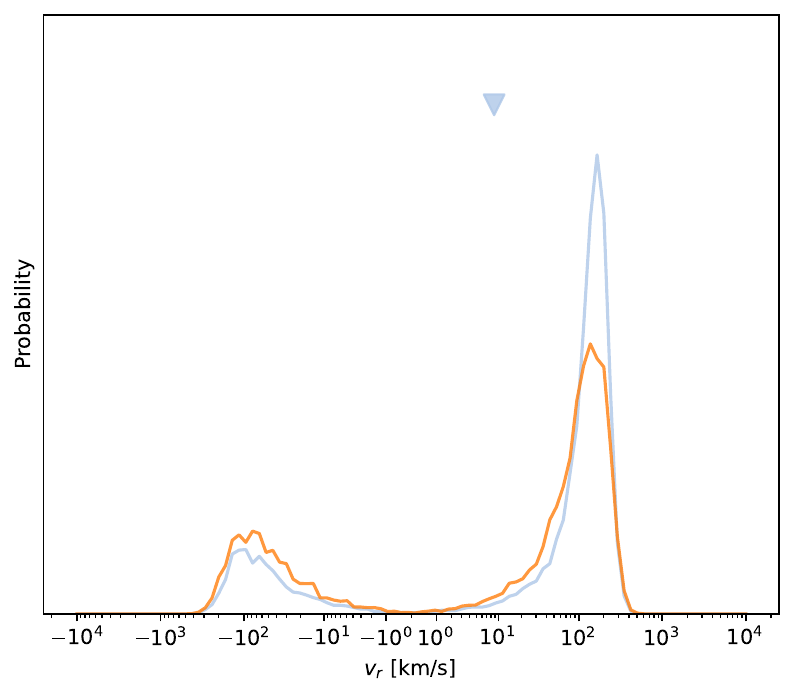}
\end{minipage}
\hfill
\begin{minipage}{0.23\textwidth}
    \centering
    \includegraphics[width=\linewidth]{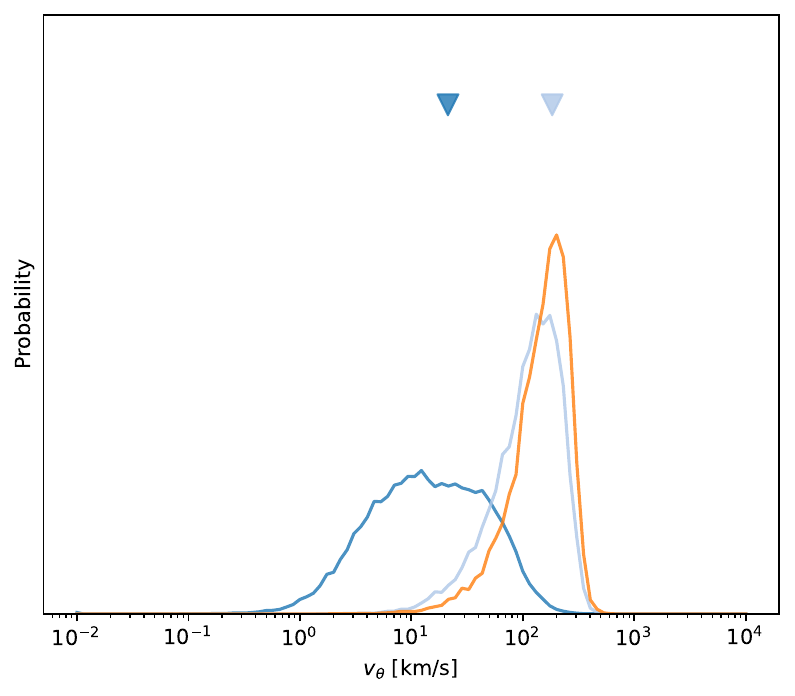}
\end{minipage}

\begin{minipage}{0.255\textwidth}
    \centering
    \includegraphics[width=\linewidth]{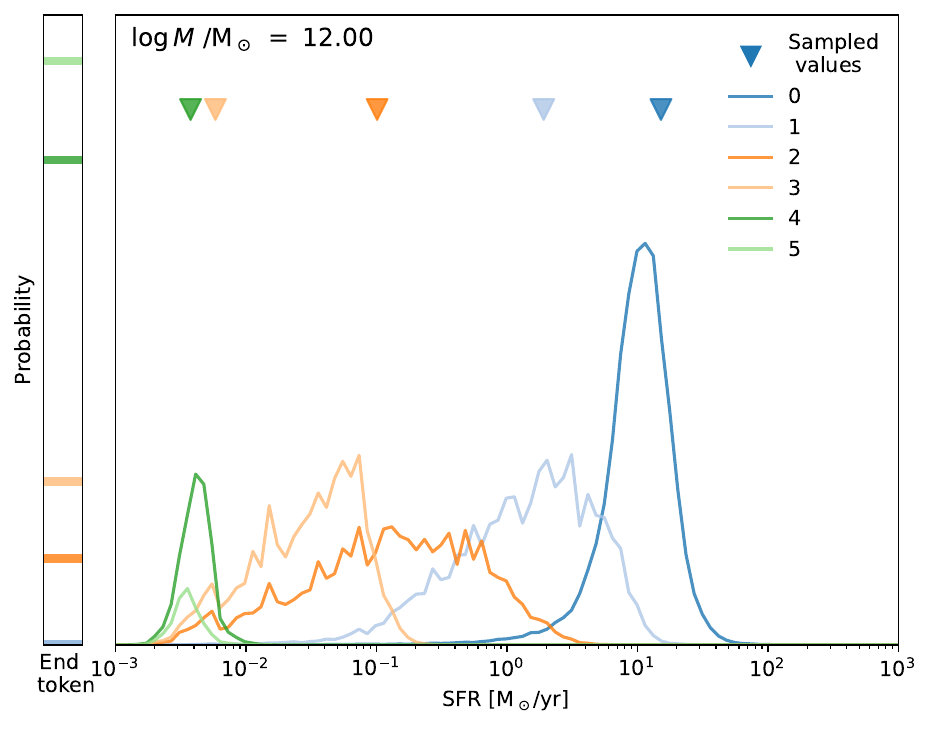}
\end{minipage}
\hfill
\begin{minipage}{0.23\textwidth}
    \centering
    \includegraphics[width=\linewidth]{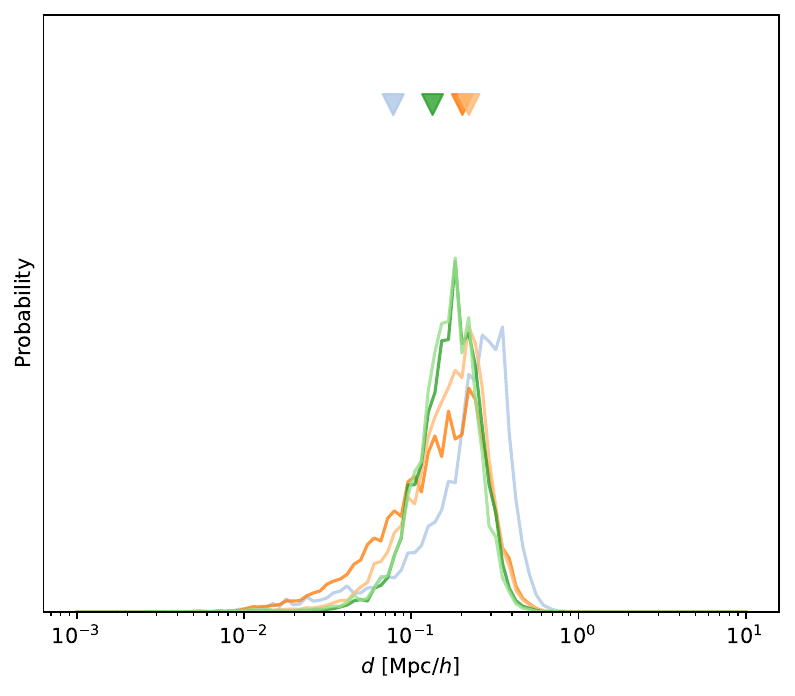}
\end{minipage}
\hfill
\begin{minipage}{0.23\textwidth}
    \centering
    \includegraphics[width=\linewidth]{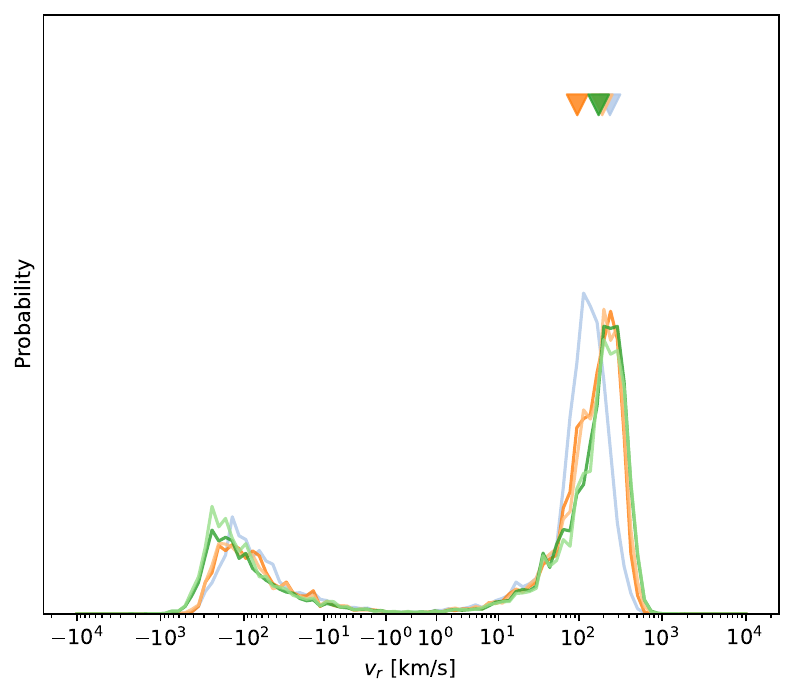}
\end{minipage}
\hfill
\begin{minipage}{0.23\textwidth}
    \centering
    \includegraphics[width=\linewidth]{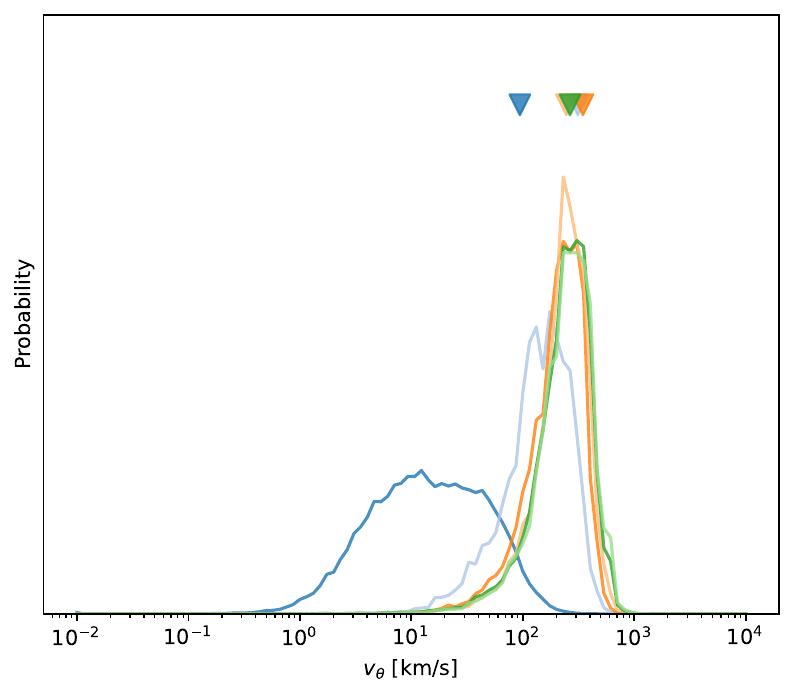}
\end{minipage}
\caption{
Two examples of the predicted probability distributions for haloes with $M = 10^{12} ~\rm M_\odot$. The inverted triangles indicate the sampled values, which are used to predict the probability distributions of the subsequent galaxies.
}
\label{fig:prob_generated}
\end{figure*}

While \cref{fig:prob_true} and \cref{fig:prob_true_d_vr_vt} show the results with true galaxy properties, 
\cref{fig:prob_generated} shows the probabilities obtained by auto-regressively feeding the generated galaxy properties as inputs.
The inverted triangles indicate the sampled values, which are then used to predict the probability distributions of the subsequent galaxies. 
The results in \cref{fig:prob_generated} demonstrate that the galaxies are sampled appropriately, i.e., the auto-regressive generation does not introduce artifacts or instabilities, even though the generated galaxies were not used as inputs during training.

\subsection{Sampled parameter distributions}
\label{app:joint_prob}

\begin{figure*}
    \centering
    \includegraphics[width=15cm]{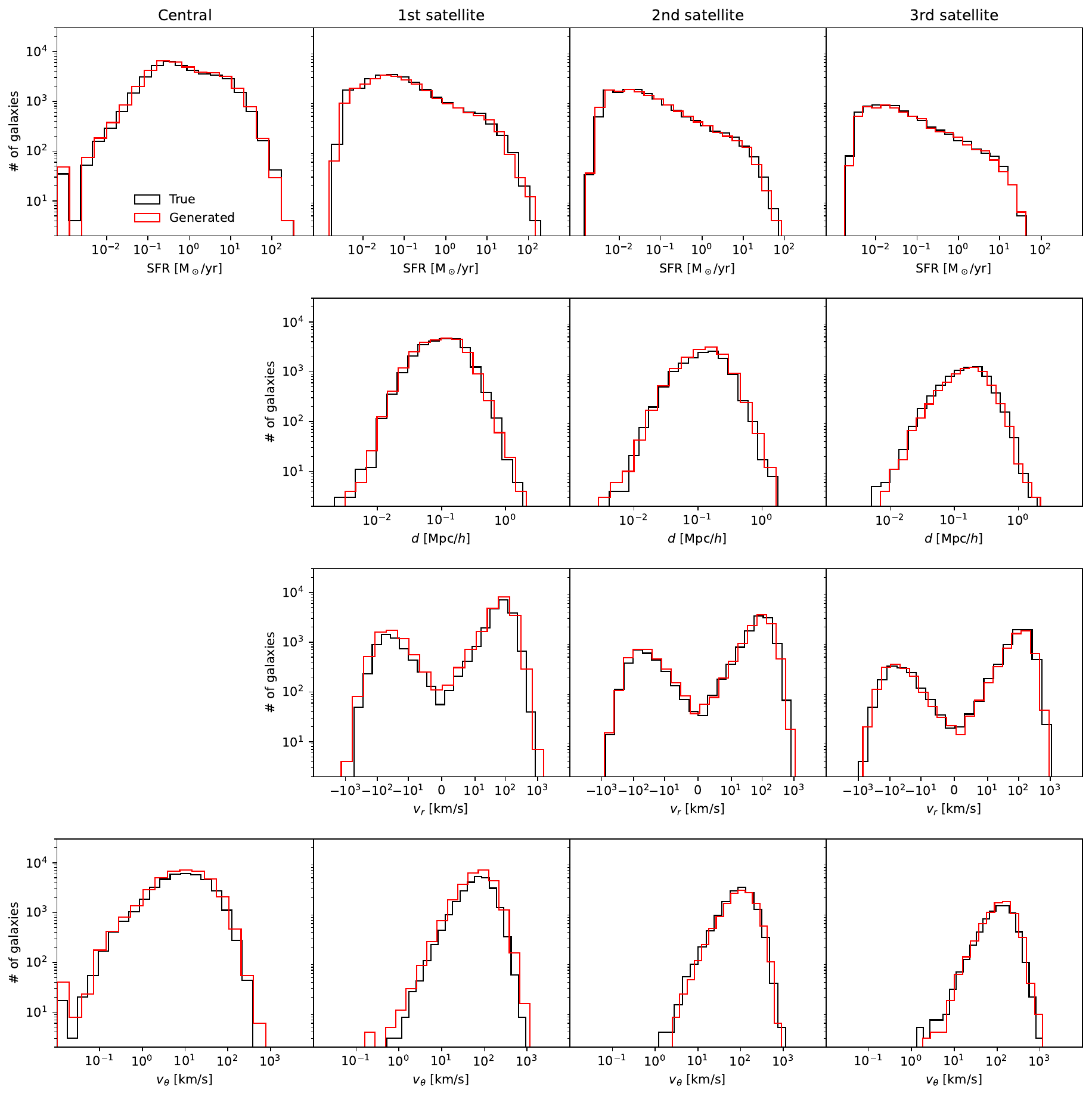}
    \label{fig:hist_x}
    \caption{
    Histograms of galaxy properties (SFR, $d$, $v_r$, and $v_\theta$ from top to bottom) in the true (black) and generated (red) datasets. From left to right, the central galaxy and the first through third satellites are shown.
    }
\end{figure*}

Figure \ref{fig:hist_x} shows histograms of SFR, $d$, $v_r$, and $v_\theta$ for the test TNG galaxies (black) and for data generated from the corresponding test TNG haloes (red). The results are presented by the rank in haloes up to the third satellite. We find that the overall marginal parameter distributions are well reproduced. We emphasise that such population-level agreement is enabled by our probabilistic generative approach. A deterministic model tends to produce distributions narrower than reality when there are intrinsically variable factors that cannot be fully described by the halo mass.

\begin{figure*}
    \centering
    \includegraphics[width=17cm]{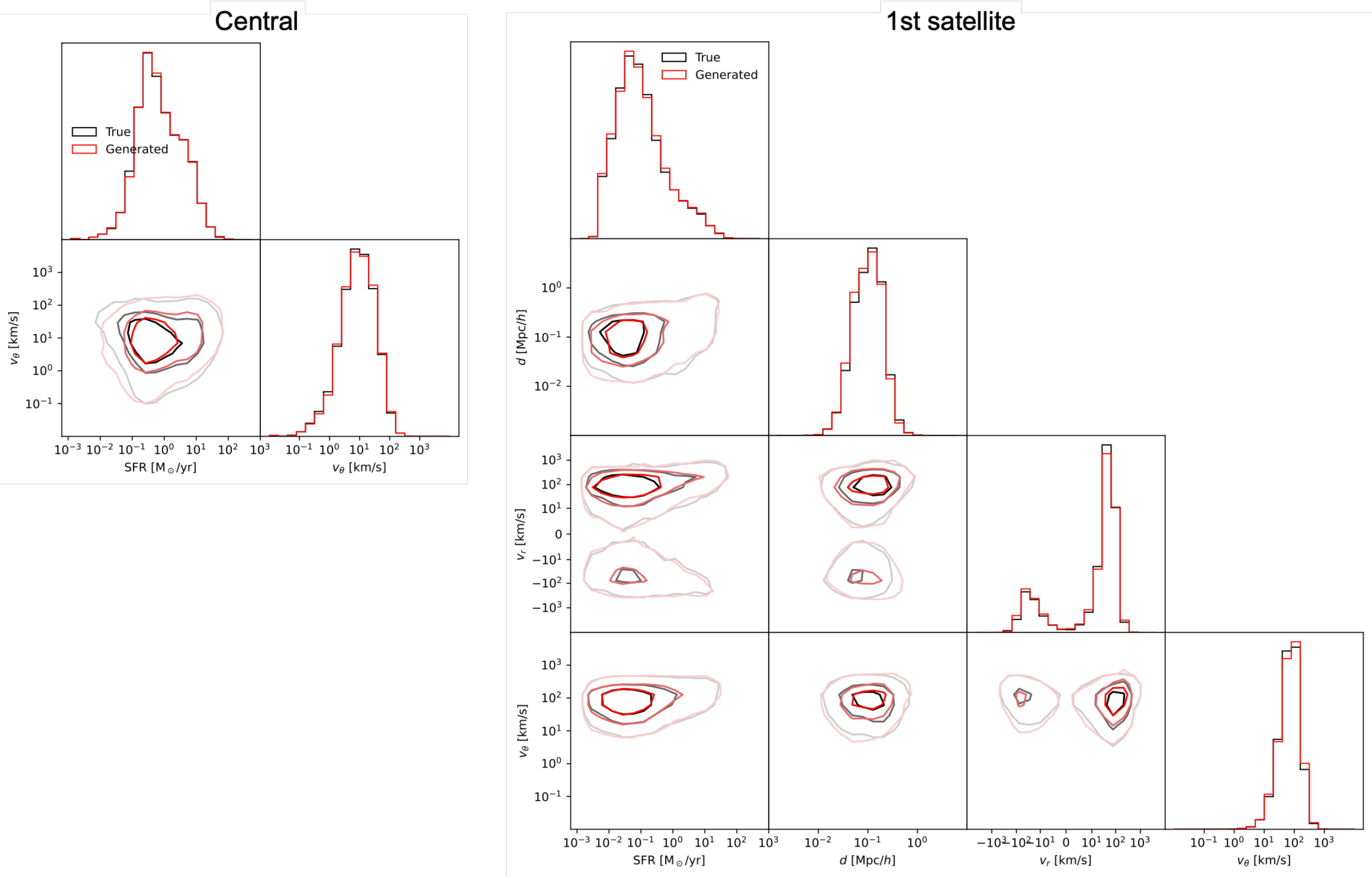}
    \caption{
    Marginal and pairwise joint distributions for the central (left) and 1st satellite (right) galaxies. The red lines show the distributions of galaxies in the test haloes, while the black lines show those of galaxies generated by our model from the same haloes.
    }
    \label{fig:joint_prob}
\end{figure*}

In our model, the correlations between parameters are not incorporated. This can lead to inaccurate joint distributions, especially when strong correlations exist among the parameters \citep[see e.g.,][]{Rodrigues23}. To examine the reproducibility of the joint distributions, we compare in \Cref{fig:joint_prob} the pairwise joint distributions of true test galaxies (black) and galaxies generated by our model from the corresponding test haloes (red). 
The results are shown for the central (left) and first satellite galaxies (right).
For the central galaxies, since the distance to the halo centre and the radial velocity relative to the halo are fixed to zero, only SFR and tangential velocity are shown.

Although our model treats the parameters independently, the overall distributions are well reproduced. This is likely because the parameters exhibit only weak correlation when the halo mass is fixed. The correlations seen in \cref{fig:joint_prob} are therefore considered spurious, arising from their dependence on halo mass. This result supports the robustness of our simplification, at least for the parameters considered in this study.

\subsection{TARP Tests}
\label{app:tarp}

We further evaluate the fidelity of our generative model using the ``Tests of Accuracy with Random Points'' \citep[TARP,][]{Lemos23}.
TARP allows us to assess whether the probability density is properly predicted without the need to explicitly compute the probabilities in high-dimensional space, relying only on true and generated samples. 
A brief overview on TARP is provided below, and we refer the readers to \citet{Lemos23} for further details.

Let $(\bm{x}^*, \bm{\theta}^*)$ denote a set of true input and target parameters, where $\bm{\theta}^*$ and $\bm{x}^*$ are sampled from the true probability distributions, $p(\bm{\theta}|\bm{x}^*)$ and $p(\bm{x})$.
Also let $\hat{p}$ denote a posterior estimator.
Suppose we can define a region in parameter space $\mathcal{D}(\alpha, \bm{x})$ with a given credible level $1-\alpha$ such that 
\begin{align} \label{eq:credible_level}
    1 - \alpha = \int_{\mathcal{D}(\alpha,\bm{x})} \hat{p}(\bm{\theta}|\bm{x}) ~{\rm d}\bm{\theta}. 
\end{align}
The expected coverage probability (ECP) of credible level $1-\alpha$ is then defined as
\begin{align}
    {\rm ECP}(\alpha) = \int_{p(\bm{x})} \int_{\mathcal{D}(\alpha,\bm{x})} p(\bm{\theta}|\bm{x}) ~{\rm d}\bm{\theta}{\rm d}\bm{x} .
\end{align}
If the posterior estimator is accurate, i.e., $\hat{p}(\bm{\theta}|\bm{x}) = p(\bm{\theta}|\bm{x})$, we have ${\rm ECP} = 1-\alpha$ for all $\alpha$. 
The TARP region is defined as
\begin{align} 
    \mathcal{D}_{\bm{\theta}_r}(\alpha,\bm{x}) = \{\bm{\theta}|d(\bm{\theta},\bm{\theta}_r) \leq R(\alpha, \bm{x})\},
\end{align}
where $\bm{\theta}_r$ is a reference parameter, $d$ is a distance metric, and $R(\alpha, \bm{x})$ is chosen such that \cref{eq:credible_level} is satisfied.
Under such a definition, \citet{Lemos23} showed that if ${\rm ECP}$ is equal to $1-\alpha$ for all $\bm{\theta}_r$, $\bm{x}$, and $\alpha$, and if both $p(\cdot |\bm{x})$ and $\hat{p}(\cdot | \bm{x})$ are continuous, then $\hat{p}(\bm{\theta}|\bm{x}) = p(\bm{\theta}|\bm{x})$. They further demonstrated that ${\rm ECP}$ deviates from $1-\alpha$ when the estimator is overconfident, underconfident, or biased.

In practice, $\rm ECP$ can be estimated using true samples, $\{\bm{\theta}^*_i, \bm{x}^*_i\}_{i=1}^{N_{\rm sim}}$, reference parameters, $\{\bm{\theta}_{r,i}\}_{i=1}^{N_{\rm sim}}$, and randomly sampled parameters by the estimator, $\{\bm{\theta}_{ij}\}_{j=1}^n$, as \citep[see][for derivation]{Lemos23}
\begin{align}
    {\rm ECP}(\alpha) = \frac{1}{N_{\rm sim}} \sum_{i=1}^{N_{\rm sim}} \mathbf{1}(f_i < 1-\alpha),
\end{align}
where
\begin{align}
    f_i = \frac{1}{n} \sum_{j=1}^n \mathbf{1} [d(\bm{\theta}_{ij}, \bm{\theta}_{r,i}) < d(\bm{\theta}^*_i, \bm{\theta}_{r,i})].
\end{align}
We compute $\rm ECP$ using the entire test dataset ($N_{\rm sim} = 38746$) and $n = 500$. The impact of sampling fluctuations on the ECP curve is expected to be minimal for the large sample size. The reference parameters are randomly sampled from $p(\bm{\theta}) \equiv \int p(\bm{\theta}|\bm{x})p(\bm{x})~{\rm d}\bm{x}$ as in the toy-model test in \citet{Lemos23}. 

\begin{figure}
    \centering
    \includegraphics[width=8cm]{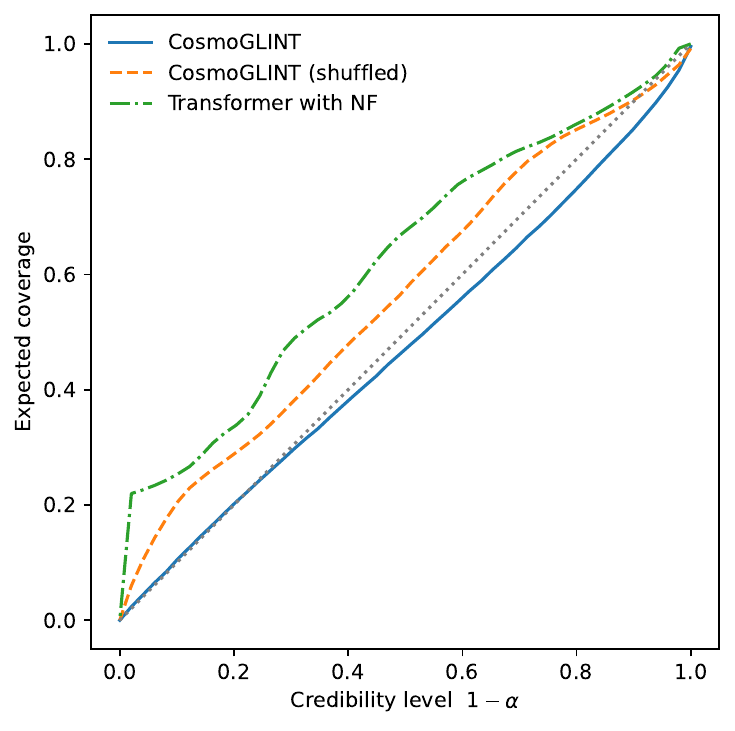}
    \caption{
    Expected coverage probabilities (ECP) as a function of credibility level. The solid, dashed, and dash-dotted lines present the results for the fiducial CosmoGLINT architecture, the shuffled dataset (see text), and the Transformer with NF (see \cref{app:nf}), respectively.
    The dotted line indicates ${\rm ECP} = 1 - \alpha$.
    }
    \label{fig:tarp}
\end{figure}

In \cref{fig:tarp}, the solid line shows the expected coverage as a function of credibility level for our model. For comparison, we constructed a shuffled dataset from the generated samples by randomizing the assignments across the $n$ realizations. In this shuffled dataset, while the overall galaxy population at a given halo mass is preserved, the correlations among galaxies within each halo are destroyed. The result with the shuffled dataset is shown as the dashed line in \cref{fig:tarp}. The shuffled case exhibits ${\rm ECP} > 1-\alpha$, indicating that the model is underconfident -- an expected behaviour given that correlations between galaxies are not taken into account. In contrast, the fiducial model (solid) lies close to the diagonal line (dotted), 
indicating that the distribution of generated samples more closely resembles that of the true samples. 
A slight tendency of ECP falling below $1-\alpha$ can be seen, which may indicate mild overconfidence or a small bias in the model. Nevertheless, the deviation is minor, suggesting that the overall performance of the model is satisfactory.

\section{Transformer with Normalizing Flow}
\label{app:nf}

In the main text, we showed the results based on a simple binning model, in which the probabilities of galaxy properties were predicted independently. A model with normalizing flows \citep[NFs][]{Rezende15,Kingma16} would, in principle, allow us to model the full joint probability distribution.\footnote{Other approaches for efficiently modelling joint probability distributions include the use of Voronoi cell-based binning in parameter space, rather than uniform binning \citep{Rodrigues25}.}
Here, we describe the NF-based model we attempted.

NF is a class of generative models that is particularly well suited for low-dimensional data and therefore considered appropriate for our task.
Several studies have employed NFs to generate galaxy populations \citep[e.g,][]{Ramanah20,Nguyen24}.
An NF constructs complex probability distributions by applying a sequence of invertible transformations to a simple base distribution.
Each transformation is bijective and differentiable, enabling both efficient sampling and exact likelihood evaluation.
In the conditional case, for an array of conditions $h$, 
the probability density of $\bm{\theta}$ is computed as
\begin{align}
    \log p(\bm{\theta} | h) = \log \pi(f_{h}(\bm{\theta})) + \log \left| \det \left( \frac{\partial f_h(\bm{\theta})}{\partial \bm{\theta}} \right) \right|, 
\end{align}
where $f_h$ is an invertible transformation that maps data $\bm{\theta}$ to latent variables, and $\pi$ denotes the distribution of the latent variables.

\begin{figure}
    \centering
    \includegraphics[width=8.5cm]{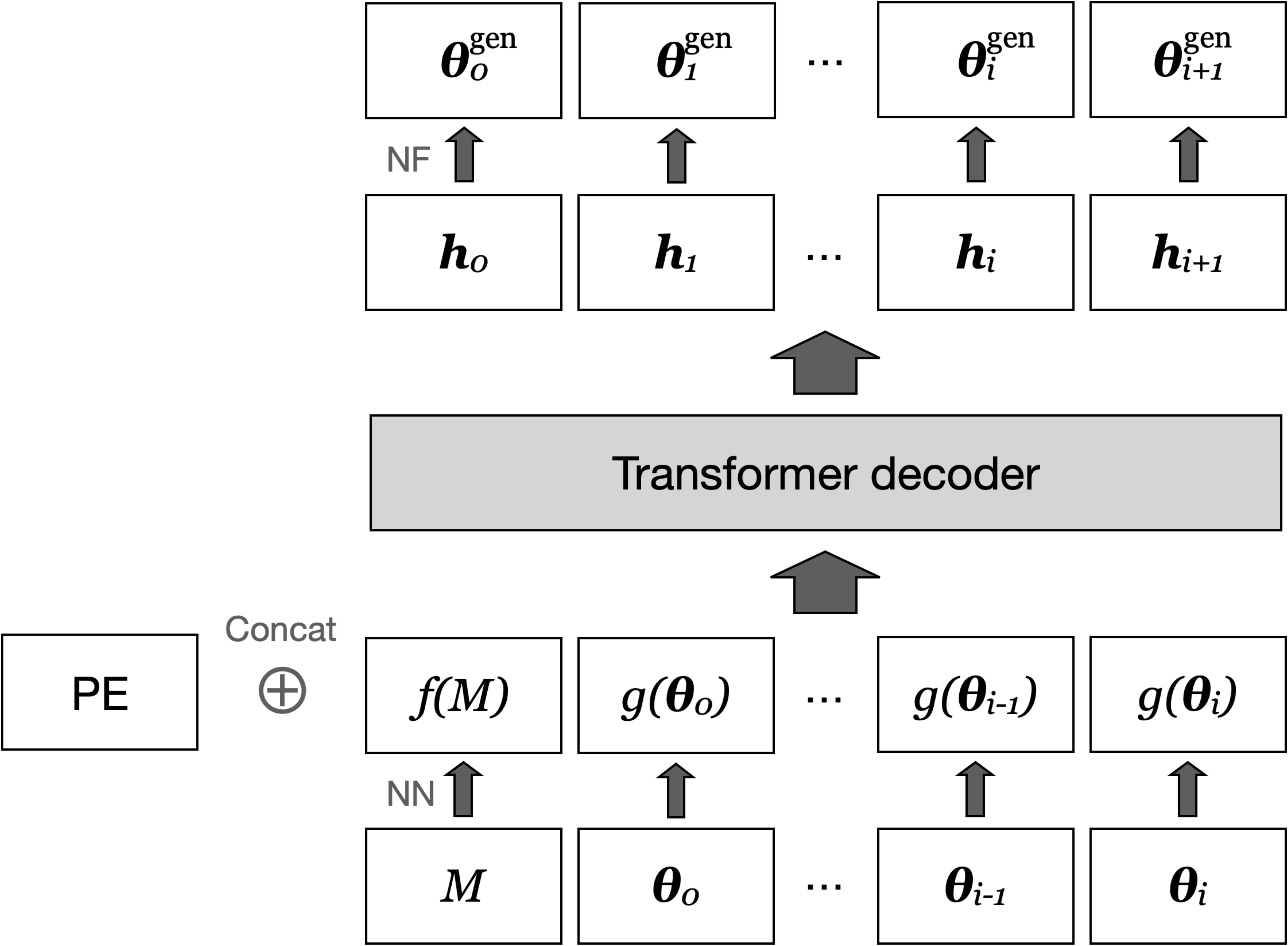}
    \caption{
    Model architecture with NF. The Transformer output is used as a conditioning input to NF.
    }
    \label{fig:architecture_NF}
\end{figure}

Our NF-based architecture is illustrated in \cref{fig:architecture_NF}.
The NF transforms latent variables $\bm{z}$ (not shown in the figure) into the galaxy properties, $\bm{\theta} = f_h^{-1}(\bm{z})$, conditioned on the Transformer output $h$, whose dimension is set to 16. Apart from the output dimension, the Transformer architecture is identical to that used in the main text. 
The NF consists of a stack of eight transformations, each consisting of (i) an affine coupling layer \citep{Dinh16} with an alternating even/odd binary mask, (ii) a piecewise rational-quadratic spline transform \citep{Durkan19}, and (iii) a random permutation of features \citep{Dinh16}.
The scale and shift parameters of the affine coupling layers are produced by a context-conditioned ResidualNet with two residual blocks, ReLU activation, and hidden dimension 512. We adopt the Gaussian distribution as the base distribution.
Following \citet{Dinh16}, to ensure that the generated values remain within the valid range, we map the data defined in [0,1] to [0.05,0.95] using an affine transformation with $\alpha = 0.05$ and then apply a logit transform before passing them to the flow.
The model is trained by maximizing the log-likelihood over the training data, using the same training/validation/test split described in the main text.
The NF is implemented using {\sc nflows}\footnote{https://github.com/bayesiains/nflows} \citep{Durkan19,nflows}.

For hyperparameter tuning, we varied the learning rate, $w_{\rm min}$, and the number of epochs as in the fiducial model (see \cref{app:hyperparameter}), as well as the number of flows and the Transformer output dimension. All tested models were able to reproduce the general parameter distributions. However, within the explored ranges, they all produced outliers with excessively large or small values with non-negligible frequencies. In the fiducial model, the occurrence of such rare outliers is restricted (see \cref{sec:model}), but in the NF-based model, it is not straightforward to impose similar constraints. Although interesting, exploring such solutions is beyond the scope of the present study.
In addition, we often encountered instabilities during the training and found that the NF models require a factor longer sampling time.

The TARP result for the NF-based model is shown as the dash-dotted line in \cref{fig:tarp}. The NF result deviates substantially from the diagonal line (dotted), with a discrepancy even larger than that of the shuffled case, indicating that the generated population is much more different from the true data. We note that this is presumably due to insufficient fine-tuning or even inappropriate model selection, and does not necessarily imply that NF is unsuitable for such tasks. Nevertheless, for ease of use, we adopt the simple binning strategy rather than the NF model in this study.

\bsp	% typesetting comment
\label{lastpage}
\end{document}